\documentclass[aps,twocolumn,prb,amsmath,amssymb]{revtex4-2}

\usepackage{amsmath}
\usepackage{amssymb}
\usepackage[varg]{txfonts}
\usepackage{color}
\usepackage[colorlinks]{hyperref}
\usepackage{tikz}
\usetikzlibrary{shapes}
\usepackage{xcolor}

\usepackage{graphicx}
\usepackage{bm}

\makeatletter
\newcommand\bigDiamond{\mathop{\mathpalette\bigDi@mond\relax}}
\newcommand\bigDi@mond[2]{\vcenter{\hbox{\m@th \scalebox{\ifx#1\displaystyle 2\else1.2\fi}{$#1\Diamond$}}}}
\newcommand{\RNum}[1]{\uppercase\expandafter{\romannumeral #1\relax}}

\def\XXint#1#2#3{{\setbox0=\hbox{$#1{#2#3}{\int}$}
		\vcenter{\hbox{$#2#3$}}\kern-.5\wd0}}

\def\be{\begin{equation}}
\def\ee{\end{equation}}

\def\bi{\begin{itemize}}
	\def\ei{\end{itemize}}
\def\bn{\begin{enumerate}}
	\def\en{\end{enumerate}}
\def\bea{\begin{eqnarray}}
\def\eea{\end{eqnarray}}
\newcommand{\bpm}{\begin{pmatrix}}
	\newcommand{\epm}{\end{pmatrix}}

\def\ba{\begin{array}}
	\def\ea{\end{array}}
\def\bd{\begin{displaymath}}
\def\ed{\end{displaymath}}

\renewcommand{\imath}{\hspace{1pt}\mathrm{i}\hspace{1pt}}

\renewcommand{\vec}{\mathbf}
\renewcommand{\Re}{\mathop{\mathrm{Re}}\nolimits}

\begin{document}
	
	\title{Optical drive of amplitude and phase modes in excitonic insulators}
	
	\author{Elahe Davari}
	\affiliation{Department of Physics, Sharif University of Technology, Tehran 14588-89694, Iran}
	\author{S. Samaneh Ataei}
	
	\affiliation{Department of Physics, Sharif University of Technology, Tehran 14588-89694, Iran}
	\author{Mehdi Kargarian}
	\email{kargarian@sharif.edu}
	\affiliation{Department of Physics, Sharif University of Technology, Tehran 14588-89694, Iran}
	\date{\today}
	
\begin{abstract}
Motivated by recent interests in exploring excitonic condensate as the ground state of some narrow-bandgap semiconductors such as transition metal dichalcogenides and layered chalcogenide material Ta$_2$NiSe$_5$, in this work we theoretically study the dynamics of condensate in response to periodically driven laser fields with different polarizations and intensities. In particular, we consider laser light beams with bicircular and circular polarizations breaking the time-revesal symmetry, and linear polarization. We show that the amplitude of the condensate oscillates in time during irradiating by light with a magnitude depending on the light intensity. The dynamics survives even after the light is switched off. The phase mode however changes linearly with time for a condensate originating from purely electronic correlations. We further show that in the presence of electron-phonon coupling the linear-in-time behavior is replaced by a harmonically oscillating behavior, a manifestation of gapped phase modes due to relative band charge symmetry breaking. Furthermore, we show that the primarily electronic and primarily lattice cases corresponding to strong and weak electron-phonon coupling, respectively, reveal distinct dynamics of the condensate, an observation which can modify the optical response of an excitonic insulator by stimulating amplitude and phase modes in the former case.         
\end{abstract}

\maketitle
\section{Introduction}
Excitons are bound states of electron and hole pairs excited across the band gap of semiconductors signifying as sharp resonant peaks in optical absorption spectroscopy. In semimetals and narrow-band gap semiconductors, a coherent formation of such pairs could lead to a ground state of condensed excitons dubbed as excitonic insulator. This state has some similarities with the superconducting state, where electrons form Cooper pairs and undergo a quantum phase transition acquiring a macroscopic quantum coherence \cite{jerome1967excitonic,halperin1968possible,kohn1967excitonic,keldysh1968collective}. Both states involve the formation and condensation of bound pairs of fermions which break the symmetry of the original system, though, there are differences between excitons and Cooper pairs, such as charge, size, binding energy, coherence length, etc., \cite{jerome1967excitonic,combescot2015excitons}. These differences affect the stability and properties of the excitonic insulator and superconductor phases leading to distinct experimental signatures. 
	
Though being introduced decades ago \cite{jerome1967excitonic,halperin1968possible,kohn1967excitonic,keldysh1968collective}, there has been a surge of interest in excitonic insulators in recent years, partly due to the vast progress in synthesizing low-dimensional systems such as graphene and transition metal dichalcogenides \cite{ma2021strongly,ataei2021evidence,sun2022evidence,varsano2020monolayer,kogar2017signatures,chernikov2014excitons,behura2021moire,wang2018colloquium,mueller2018exciton}. 
In particular, the layered dichalcogenide Ta$_2$NiSe$_5$ has been researched in  numerous experiments as an intrinsic excitonic insulator \cite{di1986physical,wakisaka2009excitonic,kaneko2012excitonic,seki2014excitonic,lu2017zero}.  
These systems offer new possibilities for creating and manipulating excitons, as well as for exploring novel phenomena such as quantum coherence, superfluidity, lasing, and quantum simulation \cite{jiang2020spin,matsuzaki2017purely,lopes2022excitonic,jiang2019half}. The control of exciton condensation is important for developing excitonic devices and applications, such as low-energy electronics, optoelectronics, and quantum information processing. However, controlling exciton condensation is not easy, and there are several challenges that need to be overcomed \cite{kogar2017signatures,mak2018opportunities,combescot2017bose,sun2017bose}. One promising route to control the exciton condensation is to use ultrafast optical pulses, which can manipulate the exciton population, coherence, and interactions on short timescales. Ultrafast optical control of exciton condensation has been demonstrated in various systems, such as transition metal dichalcogenides, organic semiconductors and polariton microcavities \cite{ball2022ultrafast,bretscher2021ultrafast,golevz2022unveiling,baldini2023spontaneous,bretscher2021imaging,chen2022optically,eroglu2020ultrafast}. However, the underlying mechanisms and dynamics of ultrafast exciton condensation have not been fully understood yet and require further investigation.

The formation of exciton condensation follows a symmetry breaking from $\mathrm{U(1)}\times \mathrm{U(1)}$ down to $\mathrm{U(1)}$ \cite{mazza2020nature,baldini2023spontaneous,zenker2014fate,sun2021second} which is described by a complex order parameter (see Sec.~\ref{sec:model and method}). The low-energy excitations are the phase mode (Goldstone mode) and the amplitude mode (Higgs mode) of the excitonic order parameter. The influence of collective modes on the optical response of the system has been studied in literatures \cite{golevz2020nonlinear, khatibi2020excitonic,murakami2020collective}. Here, however, we are interested in the following questions: how do the phase and amplitude mode respond to the time-periodic light impinging on the samples? and how do the intensity and various polarizations of the light as knobs change the response? and how does the electron-phonon coupling influence the dynamics of the collective modes?  In this paper, we aim at investigating the dynamics of the collective modes of the excitonic insulator being irradiated by a periodic light with different polarizations, in the high frequency regime where the light frequency is much larger than the band gap and the exciton binding energy. We use the Floquet theory to analyze the energy band structure of the system periodically driven by the light and to characterize the evolution of the phase mode and the amplitude mode of the excitonic insulator. We consider a simple model of spinless fermions with two orbitals sitting on the sites of a square lattice, where the onsite Coulomb interaction leads to the exciton formation. This model captures the essential physics of excitonic insulators in low-dimensional materials, such as transition metal dichalcogenides and organic semiconductors \cite{moon2021metal,thilagam2014exciton}. Our results show that the amplitude and phase of the exciton order parameter can be controlled by different light intensities and polarizations.  We show that below a critical intensity (which depends on the type of polarization), the amplitude of the exciton order parameter increases with light intensity. This indicates that the light can tune the exciton instability and so a phase transition in the system. In the absence of coupling to phonons, the phase of the exciton order parameter changes linearly with time and  could be tuned by light intensity and polarization. The coupling to phonons breaks the $\mathrm{U(1)}$ symmetry making the phase mode massive. In this case, below a critical light intensity, the phase mode oscillates with time, and if the light intensity exceeds the critical value, the phase mode changes linearly with time, which is different for different polarizations. In addition, we simulate these behavior of the phase mode of exciton with a classical model, the Kuramoto model \cite{nag2019dynamical,2016kuramoto,kuramoto1975international,acebron2005kuramoto,lotfi2018role,1988simple,moreira2019global}, which shows a good agreement with the quantum model.
	
This paper is organized as follows. In Sec.\ref{sec:model and method}, we present a model for an excitonic insulator using the mean-field treatment of the Coulomb interaction. In Sec.\ref{sec:Non-equilibrium dynamics of the system}, we investigate the non-equilibrium properties of the system under periodic light with different polarizations and employ a Kuramoto model to understand the dynamics classically. The influence of electron-phonon coupling on the dynamics is studied in Sec. \ref{sec:el-ph}.  Sec.\ref{sec:PLPE} is devoted to the study of dynamics in primarily electric and primarily lattice cases. We conclude in Sec.\ref{conclusion} and some details of Floquet theory and estimation of errors are relegated to appendices.

\begin{figure}[t]
\includegraphics[width=0.5\textwidth]{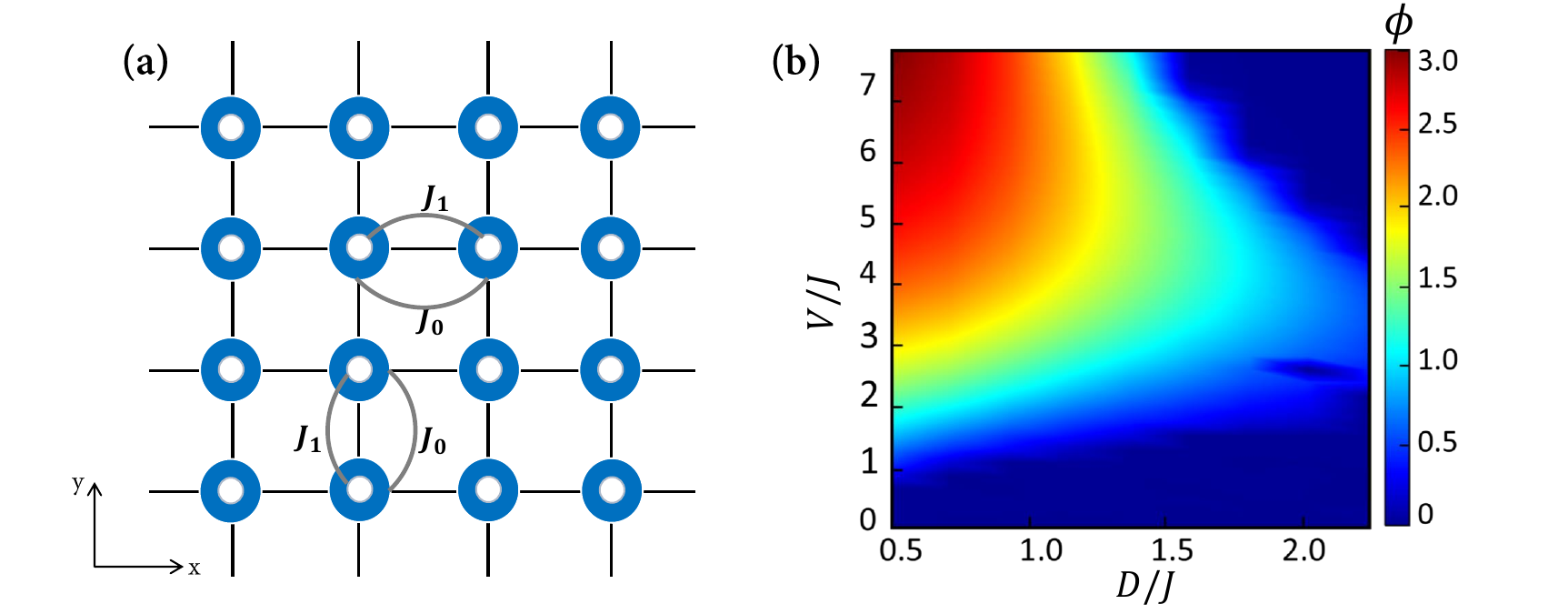}
\caption{(a) Two-dimensional square lattice with two orbitals per site. $J_{1(0)}$ are intraorbital hopping parameters between neighboring sites. (b) The equilibrium phase diagram of the model shown as density plot of the exciton order parameter $\phi$ in the plane of onsite energy $D/J$ and Coulomb interaction $V/J$.}\label{fig:lattice}
\end{figure}

\section{Model and Method} \label{sec:model and method}
We consider a two-dimensional square lattice with two orbitals per site, labeled by $\alpha=0,1$. For our purposes in this work it's enough to consider spinless electrons described by the following Hamiltonian:
	
\begin{equation}\label{eq:1}
\hat{H}=\hat{H}_0+\hat{H}_{int},
\end{equation}
where $\hat{H}_0$ is the kinetic term given by
\begin{equation}\label{eq:2}
\hat{H}_0=\sum_{\langle i, j \rangle,~\alpha} J_\alpha \hat{c}^\dagger_{i,\alpha} \hat{c}_{j,\alpha}+\sum_{i,\alpha}\left(D_\alpha-\mu\right) \hat{c}^\dagger_{i,\alpha} \hat{c}_{i,\alpha}. 
\end{equation}

Here, $\hat{c}^\dagger_{i,\alpha}$ $\left(\hat{c}_{i,\alpha}\right)$ creates (annihilates) an electron at site $i$ and in orbital $\alpha$. $J_\alpha$ is the hopping integral between $\alpha$ orbitals sitting on the neighboring sites, $D_\alpha$ is the energy level of orbital $\alpha$, and $\mu$ is the chemical potential. In momentum space 
\begin{equation}\label{eq:3}
\hat{H}_0=\sum_{k,\alpha}\left(\epsilon_{k,\alpha}-\mu\right)~\hat{c}^\dagger_{k,\alpha}\hat{c}_{k,\alpha}, 
\end{equation}
where $\epsilon_{k,\alpha}=\sum_{l=0}^3J_\alpha  e^{i\vec{k}\cdot\vec{e}_l}+D_\alpha$ is the bare electron energy dispersion with $\vec{e}_l=a(\cos \phi_l,\sin \phi_l)$,  $\phi_l=l\pi/2$ and $a$ is a lattice constant. For the sake of simplicity, we set $J_0=-J_1=J$ and $D_0=-D_1=-D$. This implies that the bands have opposite dispersion and are inverted near the $\Gamma$ point of the Brillouin zone. However, our results are not sensitive to these specific choices of parameters. We also fix the chemical potential $\mu$ such that the system is half-filled.
		
The interaction term $\hat{H}_{int}$ in \eqref{eq:1} is described by   \begin{equation}\label{eq:4}
\hat{H}_{int}=V\sum_{i}\hat{n}_{i,0}\hat{n}_{i,1}, 
\end{equation}
where  $V$ is a local interorbital Coulomb interaction and $n_{i,\alpha}=\hat{c}^\dagger_{i,\alpha}\hat{c}_{i,\alpha}$ is the electron number  operator. We treat the interaction term using the mean-field theory. By introducing 
$\phi=\langle\hat{c}^\dagger_{i,0}\hat{c}_{i,1}\rangle$
as exciton order parameter and considering the electron density in these orbitals,
$n_0=\langle\hat{c}^\dagger_{i,0}\hat{c}_{i,0}\rangle$ and 
$n_1=\langle\hat{c}^\dagger_{i,1}\hat{c}_{i,1}\rangle$,
we can rewrite the interaction \eqref{eq:4} as 
	\begin{align}
	\hat{H}_{int}^{\mathrm{MF}}=V\sum_k \left(n_1\hat{c}^\dagger_{k,0}\hat{c}_{k,0}+n_0\hat{c}^\dagger_{k,1}\hat{c}_{k,1}
	-\phi\hat{c}^\dagger_{k,1}\hat{c}_{k,0}-\phi^*\hat{c}^\dagger_{k,0}\hat{c}_{k,1}\right).  \label{eq:5}
	\end{align}
	
	Using the pseudospin $\hat{S}_k^\gamma=\frac{1}{2}\Psi_k^\dagger\sigma_\gamma\Psi_k$, where $\hat{\Psi}_k=(
	\hat{c}^\dagger_{k,0}~ \hat{c}^\dagger_{k,1})^{\mathrm{T}}$ and $\sigma_\gamma$ is the Pauli matrix for $\gamma=1-3$
	and the identity matrix for 
	$\gamma=0$, the mean-field Hamiltonian can be written in Anderson pseudo-spin representation \cite{Anderson} as $
	H^{\mathrm{MF}}=\sum_{k,\gamma} \hat{S}_k^\gamma B_k^\gamma $.  In this representation, the exciton order parameter is rewritten as 
	$\phi=\frac{1}{N}\sum_k\langle \hat{S}_k^x+i\hat{S}_k^y \rangle$ and 
	$\langle \hat{n}_0\rangle+\langle \hat{n}_1\rangle=\frac{2}{N}\sum_k \langle\hat{S}_k^0\rangle$
	and 
	$\langle \hat{n}_0\rangle-\langle \hat{n}_1 \rangle=\frac{2}{N}\sum_k \langle \hat{S}_k^z\rangle$. In addition, the components of pseudomagnetic field are computed as follows
	\begin{align}
	B^0_k&=V(n_1+n_0)\\
	B^x_k&=-2V\mathrm{Re}[\phi]\\
	B^y_k&=-2V\mathrm{Im}[\phi]\\
	B^z_k&=\epsilon_{k,0}-\epsilon_{k,1}+V(n_1-n_0)
	\end{align}
	
	Solving the mean-field equations, we map out the equilibrium ground state phase diagram as shown in Fig.~\ref{fig:lattice}(b) \cite{PWerner}.

\section{Non-equilibrium dynamics of the system} \label{sec:Non-equilibrium dynamics of the system}
The non-equilibrium dynamics of a system may reveal novel physics otherwise being  absent in equilibrium, providing a pathway to study the collective behavior of the system. Here, we present a theoretical method to study the exciton dynamics of a periodically driven system. We apply laser light with different polarizations to manipulate the symmetry of the system and investigate its effect on the evolution of the exciton order parameter. For a periodically driven system, the Floquet theory is a powerful framework for analyzing the material properties by changing the Bloch-band dispersion and the geometry of the system \cite{FloquetGoldman,meinert2016floquet,schweizer2019floquet,Floquetreview,FloquetOka,schuster2021floquet}.  In the following, we adopt the model from the previous section and simulate a half-filled excitonic system with the reference parameters $D/J=0.9$ and $V/J=4$, as an equilibrium phase of the system. By solving the self-consistent equation, we find that the exciton order parameter $\phi=0.23$ is real and opens a gap near the $\Gamma$ point of the Brillouin zone. We use the hopping parameter $J$ as our unit of energy. To express the parameter of our model in terms of $\mathrm{eV}$, we set $J=0.1\mathrm{eV}$. We use the Floquet theory to examine how the band structure is altered by the external perturbation. We then explore how the exciton order parameter evolves in the non-equilibrium state.

\begin{figure*}[t]
\includegraphics[width=0.85\textwidth]{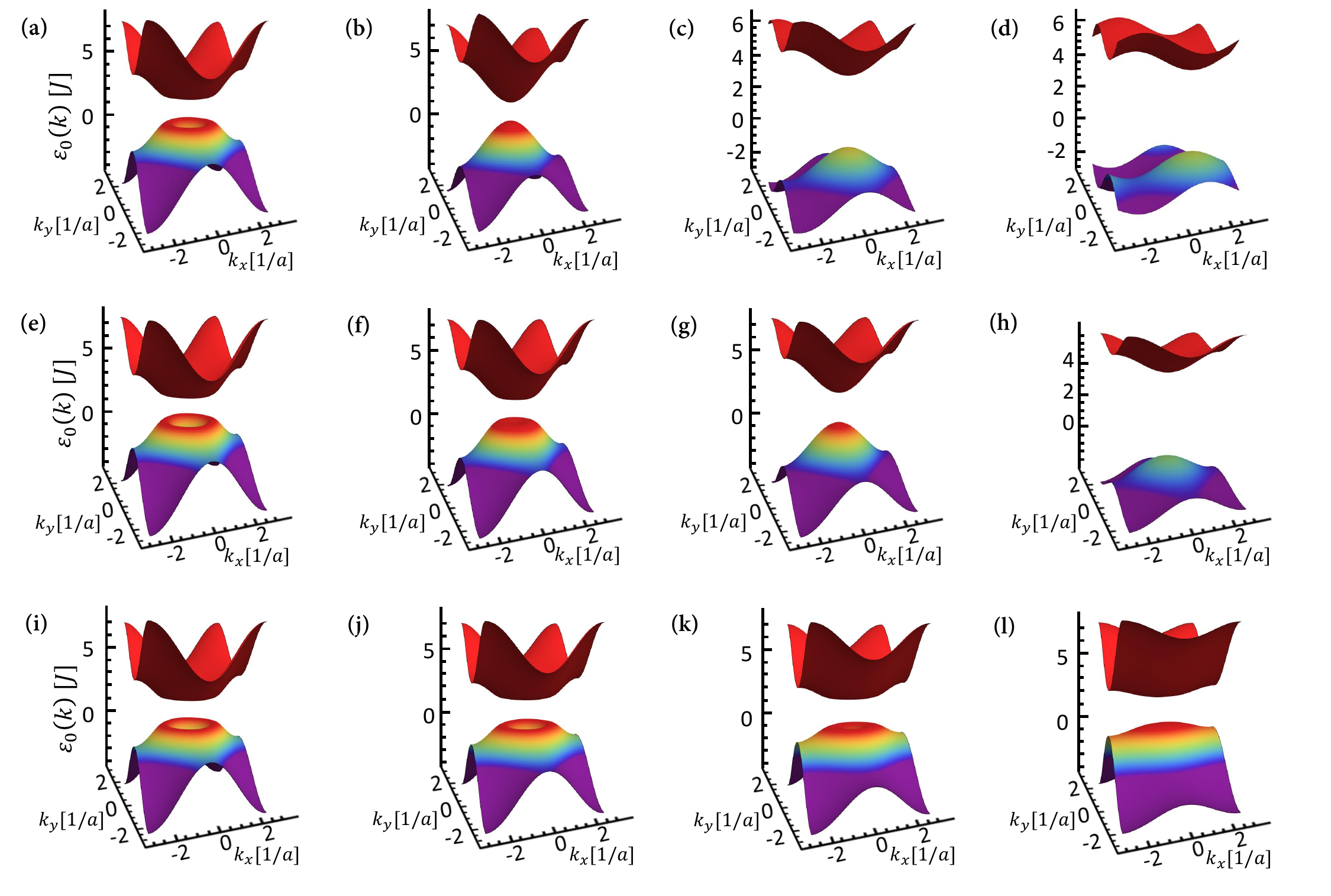}
\caption{Quasienergy spectrum of the effective Hamiltonian in the zeroth-Floquet sector for different light polarizations and intensities in the high-frequency regime ($\Omega=0.3 \mathrm{eV}$). The system is initially prepared in a regime with $D/J=0.9$ and $V/J=4$ where the exciton order parameter is $\phi=0.23$. We set the hopping magnitude $J=0.1 \mathrm{eV}$ and the lattice constant $a=3.5 \mathring{A}$. The figure shows the quasienergy variation for (a-d) 3-fold BCL with $(m_1,m_2)=(1,-2)$, $\theta=\pi/3$ and  $A_1=A_2=A$, (e-h) CL with $(m_1,m_2)=(1,0)$, $A_1=A$ and $A_2=0$ and  (i-l) linearly polarized light along the $x$ direction with $(m_1,m_2)=(1,-1)$, $\theta=0$ and  $A_1=A_2=A/2$. The light intensity ranges from $A=1.0~ \mathrm{\mu V.s.m^{-1}} $ to $A=4.0~\mathrm{\mu V.s.m^{-1}}$ with a step of unity from left to right. The symmetry breaking and the gap renormalization are more pronounced at high intensities.} 
\label{FB}
\end{figure*}
	
\begin{figure*}[t]
\includegraphics[width=0.7\textwidth]{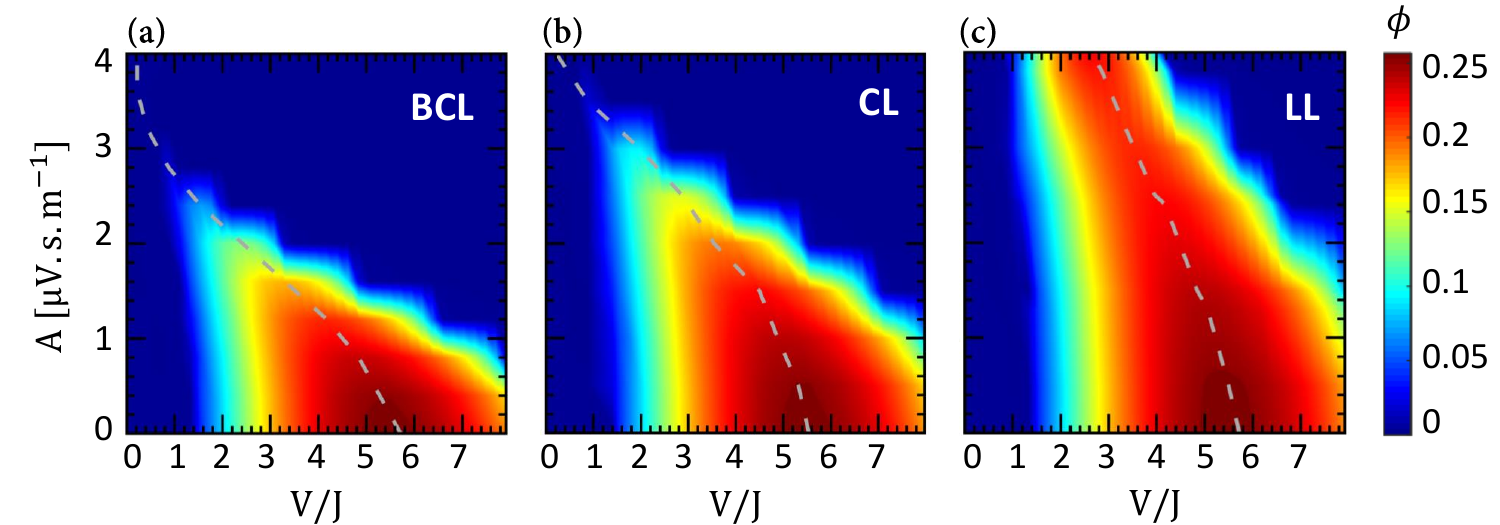}
\caption{The effect of light intensity $A$ and interaction strength $V/J$ on the exciton order parameter $\phi$ for different light polarizations: (a) BCL  with $(m_1,m_2)=(1,-2)$, $\theta=\pi/3$ and  $A_1=A_2=A$, (b) CL with $(m_1,m_2)=(1,0)$, $A_1=A$ and $A_2=0$, and (c) linearly polarized light along the $x$ direction with $(m_1,m_2)=(1,-1)$, $\theta=0$ and  $A_1=A_2=A/2$. The dashed gray line indicates the BCS-BEC crossover in the excitonic phase. The on-site energy is $D/J=0.9$.} 
\label{phaseOPAV}
\end{figure*}

\subsection{The Floquet band structure}
In this subsection, the Floquet band structure of the two-dimensional square lattice is calculated using the Floquet Hamiltonian (see Appendix \ref{APP_A}). Coupling to the electric field of an incident light, the mean-field Hamiltonian reads as
\begin{align}
\hat{H}^{\mathrm{MF}}(t)=\frac{1}{N}\sum_k
\hat{\Psi}_k^\dagger
\hat{H}^{\mathrm{MF}}\left(\vec{k}+\frac{e}{\hbar}\vec{A}(t)\right)\hat{\Psi}_k, \label{eq:14}
\end{align}
where the electric field of light is incorporated into this model via Peierls substitution with the vector potential $\vec{A}(t)$. $e$ and $\hbar$ are the electron charge and the reduced Planck constant, respectively. The function describing
$\vec{A}(t)$ depends on the polarization of the light, generally defined as
\begin{align}
A(t)=A_1 e^{im_1\Omega t} + A_2 e^{im_2\Omega t+i\theta}, \label{eq:15}
\end{align}
which consists of two circularly polarized lights with amplitudes $A_1$ and $A_2$
and different harmonics $m_1$ and $m_2$. The $\Omega$ is the frequency and the  parameter $\theta$ indicates the phase difference between these two circular components. With this choice of the vector potential, the time dependent Hamiltonian $\hat{H}^{\mathrm{MF}}\left(\vec{k}+\frac{e}{\hbar}\vec{A}(t)\right)$ can be written as
\begin{align}
\hat{H}^{\mathrm{MF}}\left(\vec{k}+\frac{e}{\hbar}\vec{A}(t)\right)&=\sum_{\alpha} \bigg(\sum_{l=0}^3J^l_\alpha(t)  e^{i\vec{k}\cdot\vec{e}_l} +D_\alpha+V n_{\bar{\alpha}}\bigg)\hat{c}_{k,\alpha}^\dagger\hat{c}_{k,\alpha}\nonumber\\
&-V\phi^*\hat{c}^\dagger_{k,0}\hat{c}_{k,1}+\mathrm{H.c.}, \label{eq:16}
\end{align}
where $J^l_\alpha(t)=J_\alpha e^{i\frac{e}{\hbar}a A_1\cos(m_1\Omega t-\phi_l)}e^{-i\frac{e}{\hbar} a A_2\cos(m_2\Omega t+\theta-\phi_l)}$. Fourier transformed to frequency domain (see Eq.\eqref{eq:A3}), the Floquet Hamiltonian is obtained as 
\begin{align}
\hat{H}_F=
\begin{pmatrix}
	\ddots& & \vdots & \vdots & \vdots& &\\
	\cdots&&\hat{H}_0-\Omega & \hat{H}_1 &  && \\
	&& \hat{H}_{-1}& \hat{H}_0 & \hat{H}_1 & & \\
	& && \hat{H}_{-1} &\hat{H}_0+\Omega &&\cdots \\
	& && \vdots & \vdots &&\ddots \\
	\\
	\end{pmatrix}, \label{eq:17}
	\end{align}
	where $\hat{H}_n$ are the Fourier components of the mean-field Hamiltonian and is given by
	\begin{align}
	\hat{H}_n(k)&=\sum_{\alpha} \bigg(\sum_{l=0}^3J_\alpha^{nl}e^{i\vec{k}\cdot\vec{e}_l}+\big(D_\alpha+ Vn_{\bar{\alpha}}\big)\delta_{n,0}\bigg)\hat{c}_{k,\alpha}^\dagger\hat{c}_{k,\alpha}\nonumber\\
	&-\bigg(V\phi^*\delta_{n,0}\bigg)\hat{c}^\dagger_{k,0}\hat{c}_{k,1}+\mathrm{H.c.},\label{eq:18}
	\end{align}
	
	Equation \eqref{eq:18} shows that the effect of the light is to renormalize the hopping integral as 
	\begin{align}\nonumber
	J_{\alpha}^{nl}=&(1-2\alpha)Je^{in/m_1(\phi_l-\pi/2)}\\
	&\times
	\sum_mi^{m\left(\frac{m_2}{m_1}+1\right)} e^{-im\theta}e^{im\phi_{l}\left(1-\frac{m_2}{m_1}\right)}\mathcal{J}_{\frac{(-n+m m_2)}{m_1}}\left(\frac{e}{\hbar} a A_1\right)\mathcal{J}_m\left(\frac{e}{\hbar} a A_2\right),\label{eq:19}
	\end{align} 
where $\mathcal{J}_n$ is the $n$-th Bessel function. As an approximation for numerical calculation of the Floquet spectrum, we focus on the high-frequency limit
$\Omega>t_0,V$. In this limit, Floquet sectors in Eq.\eqref{eq:17} split in energy and one can restrict to the zeroth-Floquet sector. Thus, the Floquet Hamiltonian will be replaced by a two-dimensional effective Hamiltonian which is equivalent to its time average \cite{mikami2016brillouin}.
	
In the following, by adjusting the incident light parameter, we will investigate the Floquet Hamiltonian and the evolution of energy band structure for different polarizations: bicircular, circular, and linear.
	
	\textbf{Bicircular light} - Bicircular light (BCL) polarization is a superposition of two circularly polarized lights (CL) with opposite chirality and different frequencies. According to Eq.\eqref{eq:15}, the vector potential of the BCL possesses a rose pattern so that the integer numbers $m_1$ and $m_2$ control the number of its leaves with $(m_1+m_2)/\mathrm{gcd}(m_1,m_2)$-fold rotational symmetry. The phase difference $\theta$ between these two CLs can rotate the shape of this rose pattern or equivalently change the direction of electric polarization. Due to the spatial pattern of the electric field of BCL, it can break not only time-reversal symmetry but also inversion symmetry of the system \cite{trevisan2022bicircular,nag2019dynamical}.
Using Eq.\eqref{eq:15} and choosing $(m_1,-m_2)=(1,2)$,
where $A_x(t)=A_1 \cos(\Omega t)+A_2 \cos(2\Omega t-\theta)$ and
$A_y(t)=A_1 \sin(\Omega t)-A_2 \sin(2\Omega t-\theta)$, we explore the effect of three-fold BCL on the electronic band structure. Our results show that the three-fold BCL leads to band renormalization and rotational symmetry breaking, which depend only on the light intensity ($A_1$ and $A_2$) and are independent of the polarization angle $\theta$ due to the local excitation.
	
	\textbf{Circular light} - For circular light polarization, we choose one of the circular components of the vector potential in Eq.\eqref{eq:15}, i.e., $(m_1,m_2)=(1,0)$, $\theta=0$ and $A_1=A,~A_2=0$. In this case the time-reversal symmetry is broken in the system, while the rotational symmetry is preserved.

	\textbf{Linear light} -  The linearly polarized light (LL), which is characterized by $(m_1,m_2)=(1,-1)$, $\theta=0$ and $A_1=A_2=A/2$, breaks the original discrete rotational symmetry.  We show that by tuning the parameter of incident light for a linearly polarized light along the $x$ direction, the energy dispersion in the $x$ and $y$ directions becomes asymmetrically renormalized. This effect becomes more pronounced as the intensity of the incident light increases.
	
Fig.~\ref{FB} shows the quasienergy spectrum of the zeroth-Floquet sector in the first Brillouin zone for different light polarizations. The effect of the light can be seen as a change in the bandwidth as well as the symmetry of the distribution of the condensate in momentum space, intensified as red gloming in the valence band. The change of zeroth-Floquet spectra for BCL and CL polarization with light intensity are shown in Fig.~\ref{FB}(a-d) and Fig.~\ref{FB}(e-h), respectively. By increasing the intensity, the momentum distribution moves toward the center of the Brillouin zone and retains the original symmetry of the system. However, since BCL polarization breaks the lattice symmetry, the distribution becomes asymmetric. The spectrum remains symmetric for CL polarization. The light intensity strongly renormalizes the bandwidth, flattening it and making the band gap larger than in equilibrium. These variations result from the renormalization of hopping integrals $J_0^{0l}$ and $J_1^{0l}$ in \eqref{eq:19} which contain the Bessel functions. According to Eq.\eqref{eq:19} for large values of $A$, the hopping values decrease, which leads to the flattening of the energy bands and a phase transition in the system. Fig.~\ref{FB}(i-l) depicts the spectra for LL polarization. Again, at high intensities the momentum distribution is elongated along the $k_x$ axis due to the $x$-linear polarization of the light.   
	
It is also instructive to see how the intensity of optical fields may change the phase diagram of the equilibrium system.  To do this, we calculate the excitonic order parameter for our photo-excited system by diagonalizing the zeroth Floquet sector Hamiltonian and solving the self-consistent equations. Fig.~\ref{phaseOPAV}(a-c) shows how the order parameter changes with the light intensity $A$ and the Coulomb interaction $V/J$ for different light polarizations: BCL, CL, and LL. The dashed gray line marks the BCS-BEC crossover in the excitonic phase. In the BCS condensate the pairs are mainly condensed around the minimum of the gap occurring at finite momenta, while the BEC condensate pairs concentrate around $\mathbf{k}=0$.  We see that the light intensity and polarization can induce the crossover from BCS to BEC regimes at lower interactions, where the condensate moves from finite to zero momenta. However, if the intensity is too high, the condensate will disappear.
	

\subsection{Non-equilibrium dynamics of the exciton order parameter} 
To study the non-equilibrium dynamics of the exciton condensate, we use the time-dependent mean-field Hamiltonian \eqref{eq:14} in the pseudo-spin representation. The time evolution of mean-field parameters is given by the Heisenberg equation of motion, 
\begin{align}
\frac{d\langle\vec{S}_k(t)\rangle}{dt}=~\vec{B}_k(t)\times \langle\vec{S}_k(t)\rangle. \label{eq:20}
\end{align}
	
The above equation is a set of equations that we solve using the fourth-order Runge-Kutta method, and the exciton order parameter $\phi(t)$ is calculated at each time step. Cares have to be taken to treat the accumulation of errors properly when solving the equations of motion numerically as we explain in Appendix \ref{B}. In our calculations we assume the zero temperature limit and consider the equilibrium state of the system as initial point.

\begin{figure*}[t]
\includegraphics[width=0.9\textwidth]{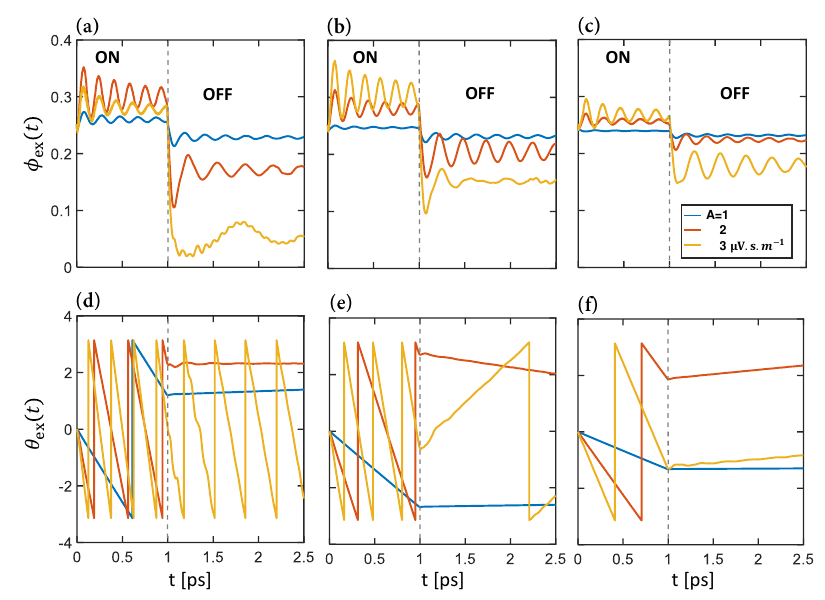}
\caption{ The evolution of the exciton order parameter for different light polarizations in the high frequency limit $\Omega=0.3~ \mathrm{eV}$. The system is initially in the BCS regime with $V/J=4$, $D/J=0.9$ and exciton mean field parameter $\phi=0.23$ as an equilibrium phase. (a-c) The collective amplitude mode of the excitonic phase under BCL, CL and LL, respectively. (d-f) The collective phase mode of the excitonic phase under BCL, CL and LL, respectively.  Note that the phase mode is plotted modulo $2\pi$, which causes the apparent zig-zag shape of the curves. The actual phase mode varies linearly in time with different rates for different light polarization and intensity. The dashed gray line marks the light cutoff time, which is chosen at $t=1.0~$ps here.} 
\label{fig.4}
\end{figure*}

	
As shown in Fig.~\ref{fig.4}, the exciton order parameter, $\phi(t)$, becomes complex when the system is driven out of equilibrium by lights with different polarizations as 
\begin{equation}
\phi(t)=\phi_{ex}(t)e^{i\theta_{ex}(t)}, \label{eq:21}
\end{equation}
where $\phi_{ex}(t)$ and $\theta_{ex}(t)$ are the amplitude and the phase of the exciton order parameter, respectively. We will explore how $\phi_{ex}(t)$ and $\theta_{ex}(t)$ depend on different light polarizations in two regions of light intensity: switch ON and switch OFF.
	
\subsubsection{Dynamics of amplitude mode}
The time evolution of the amplitude mode, $\phi_{ex}(t)$, is shown in Fig.~\ref{fig.4}(a-c). Each panel shows the evolution for a given polarization and different plots correspond to different values of light intensities $A= 1, 2, 3 ~\mathrm{\mu V.s.m^{-1}}$ (corresponding to light intensities of about 0.4, 1.2, 2.7~mJ/cm$^2$). The first observation is that in the switch-ON region, the amplitude of the exciton is influenced by both polarization and intensity of the incident light and $\phi_{ex}(t)$ oscillates coherently with a frequency that matches the minimum gap  $E_g=2V\phi/J\simeq2$. When driven by the BCL polarization, the amplitude of the oscillations first increases with light intensity and then decreases for more intensive fields. The amplitude of oscillations however is enhanced with the intensity of light for CL and LL polarizations. The different responses between polarizations could be traced back to the dynamics of the exciton order parameter in momentum space. Note that $\phi_{ex}(t)=N^{-1}\sum_{\mathbf{k}} \langle S^{x}_{\mathbf{k}}(t)+iS^{y}_{\mathbf{k}}(t)\rangle$. The stimulus optical drive with a given polarization and intensity would generate a pseudo-spin dynamics in momentum space, and hence, the collective behavior obtained by integrating over pseudo-spins depends on the distribution of the dynamics induced by the light.                

The second observation is that when the light is turned off, the oscillations of $\phi_{ex}(t)$ survive with a lower amplitude which depends on the light intensity before turning it off. Physically the mode should be damped over time, but in the absence of any dissipation in our model the coherent oscillations could survive at least for weakly dissipative system. Also, we observe that the amplitude is diminished more severely for high intensive optical fields before turning it off.

\subsubsection{Dynamics of phase mode}
The evolution of the phase of order parameter, $\theta_{ex}(t)$, is shown in Fig.~\ref{fig.4} (d-f). The phase mode changes linearly with time with a slope which  depends on the incident light polarization and intensity. Note that the plots depict $\theta_{ex}(t)~\mathrm{mod}~2\pi$, i.e., the linear-in-time behavior is folded to $[-\pi,\pi]$ interval. In particular, for BCL the slope is steeper than CL and LL polarization and increases with the light intensity. This behavior of the phase of the exciton order parameter also shows that the phase mode is affected by the symmetry of the electronic bands, which is controlled by the incident light. In other words, the distribution of the exciton order parameter in momentum space over the Brillouin zone is determined by the light properties. This leads to an increase in the slope of the time evolution of the phase of exciton diagram at high intensities. In the switch OFF region, the phase of the order parameter continues to change linearly but with a lower slope than in the ON region. It is also worth mentioning that our results from the evolution of the phase of exciton order parameter are consistent with the theoretical predictions of Golež \textit{et al.}\cite{golevz2020nonlinear}, where the authors studied the nonlinear spectroscopy of collective modes in excitonic insulators and found that without electron-phonon coupling, the equation of motion of the phase mode of the exciton order parameter is given by $\ddot{\theta}_{ex}(t)=0$, resulting in a linear-in-time behavior for $\theta_{ex}(t)$. 
	
%

\subsection{Classical description using Kuramoto model }
To understand the overall behavior of the exciton order parameter explained above, it is instructive to use a classical model based on the Kuramoto model. This model describes the dynamics of N-coupled classical oscillators \cite{nag2019dynamical,2016kuramoto,kuramoto1975international,acebron2005kuramoto,lotfi2018role,1988simple,moreira2019global}. To see this, we represent an exciton at each point of the Brillouin zone by a dipole with direction and amplitude given by the vector $(S_k^x,S_k^y)$. We assume that the components of spinor are $\psi_{k,0}=e^{i\theta_{k,0}}\cos\gamma_k$ and $\psi_{k,1}=e^{i\theta_{k,1}}\sin\gamma_k$ satisfying $|\psi_{k,0}|^2+|\psi_{k,1}|^2=1$. Then, we define the exciton order parameter as
\begin{equation}
\phi(t)=\frac{1}{N}\sum_k\frac{\sin(2\gamma_k)}{2}e^{i\theta_k}, \label{eq:22}
\end{equation}
where $\sin(2\gamma_k)/2$ and $\theta_k=\theta_{k,1}-\theta_{k,0}$ are the  amplitude and polarization angle of the exciton at each momentum, respectively. Using the Hamiltonian \eqref{eq:14} in the time-dependent Schr\"{o}dinger equation, we obtain the dynamics of an exciton order parameter as: 
	
\begin{equation}
i\frac{d}{dt}\langle \psi_{k,0}\psi_{k,1}\rangle=-\dot{\theta}_k e^{i\theta_k} \frac{\sin(2\gamma_k)}{2}-ie^{i\theta_k} \dot{\gamma}_k \cos(2\gamma_k). \label{eq:23}
\end{equation}
	
The above equation leads to the two coupled differential equations for amplitude and angle of polarization given by 
	
\begin{align}
&\dot{\theta}_k=B_k^z(t)+\cot(2\gamma_k)\left(B_k^x(t)\cos\theta_k+B_k^y(t)\sin\theta_k\right),\label{eq:24}\\
&\dot{\gamma}_k=\frac{1}{2}\bigg(B_k^x(t)\sin\theta_k-B_k^y(t)\cos\theta_k,\bigg)\label{eq:25}
\end{align} 
where 
\begin{align}
&B_k^x(t)=-\frac{V}{N}\sum_p\sin(2\gamma_p)\cos\theta_p.\label{eq:26}\\
&B_k^y(t)=-\frac{V}{N}\sum_p\sin(2\gamma_p)\sin\theta_p.\label{eq:27}
\end{align}
	
The equations \eqref{eq:24} and \eqref{eq:25} can be rewritten as
	
\begin{align}
\dot{\theta}_k=&\bigg(\epsilon_{k+\frac{e}{\hbar}A(t),0}-\epsilon_{k+\frac{e}{\hbar}A(t),1}+V(n_1-n_0)\bigg)\nonumber\\
&+\frac{V}{N}\cot(2\gamma_k)\sum_p\sin(2\gamma_p)\cos(\theta_p-\theta_k),\label{eq:28}\\
&\dot{\gamma}_k=\frac{-V}{2N}\sum_p\sin(2\gamma_p)\sin(\theta_p-\theta_k).\label{eq:29}
\end{align}
	
Now let us compare these equations with the Kuramoto model. The latter model is given by \cite{moreira2019global,2016kuramoto,kuramoto1975international,acebron2005kuramoto,lotfi2018role,1988simple}
\begin{align}
\dot{\eta}_i=\omega_i+\sum_{j=1}^N M_{ij}\sin(\eta_i-\eta_j)+\epsilon_i(t)+F \sin(\sigma t-\eta_i),\label{eq:30}
\end{align} 
which describes the synchronization of phases in a system consisting of N-coupled oscillators each with phase $\eta_i$ and individual frequency $\omega_i$. $\epsilon_i(t)$ is a noise term that oscillates very fast in time. $F$  and $\sigma$ are the strength and the frequency of the external force, respectivly. The external force can have phase and/or time dependence and may influence the frequency and the phase of the oscillators. The parameter $M_{ij}$ determines the degree of synchronization between the phases of different oscillators \cite{kuramoto1984chemical,moreira2019global}. Comparison of equations \eqref{eq:28} with \eqref{eq:30} reveals that $\theta_k$ plays the role of $\eta_i$ and $V\cot(2\gamma_k)\sin(2\gamma_p)$ corresponds to  coupling between the phases of excitons of different modes, which is analogous to $M_{ij}$ in the Kuramoto model. Also, similar to $\omega_i$, $\omega_k\equiv\left(\epsilon_{k+\frac{e}{\hbar}A(t),0}-\epsilon_{k+\frac{e}{\hbar}A(t),1}+V(n_1-n_0)\right)$ indicates the individual frequency of each mode that couples to light with vector potential $A(t)$. We expand this term and according to Eq.\eqref{eq:18}, it is approximated by 
	\begin{equation} 
	\omega_k \simeq \sum_{l=0}^3 \big(J_0^{0l}-J_1^{0l}\big)e^{i\vec{k}\cdot\vec{e}_l}+V(n_1-n_0), \label{eqomega}
	\end{equation}
where we only keep the first term of the Fourier expansion because higher terms oscillate very fast in the limit of high frequency and act as a noise term $\epsilon_i$ in \eqref{eq:30}. In addition, there is no external force, $F \sin(\sigma t-\eta_i)$ in \eqref{eq:30}, that aligns the exciton dipoles with themselves.
	
Next, we use Eq.\eqref{eq:28} to describe the evolution of the exciton order parameter in momentum space in a system driven by light. The equations \eqref{eq:21} and \eqref{eq:22} show that these microscopic evolutions in momentum space result in the amplitude and the phase of exciton as
\begin{align}
\phi_{ex}&=\frac{1}{2N}\sqrt{\left(\sum_k \sin(2\gamma_k)\cos\theta_k\right)^2+\left(\sum_k \sin(2\gamma_k)\sin\theta_k\right)^2},\label{eqamp}\\
\theta_{ex}&=\arctan\left(\frac{\sum_k \sin(2\gamma_k)\sin\theta_k}{\sum_k \sin(2\gamma_k)\cos\theta_k}\right). \label{eqphase}
\end{align}
	 
\begin{figure*}[t]
\includegraphics[width=0.85\textwidth]{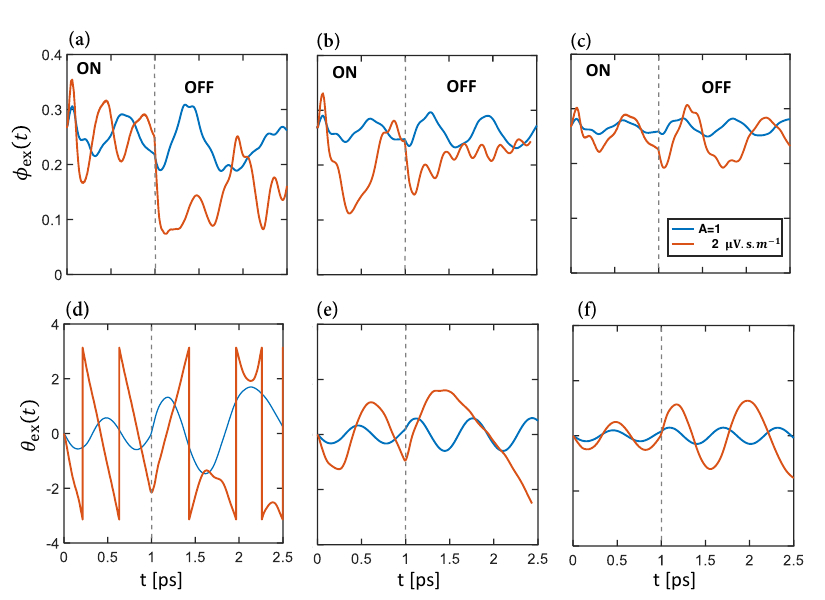}
\caption{The evolution of the exciton order parameter in the presence of the electron-phonon coupling for different light polarizations and intensities in the high frequency limit $\Omega=0.3~\mathrm{eV}$. The system is initially prepared in the BCS regime with $V/J=4$, $D/J=0.9$, $\omega_{ph}/J=0.1$ and the effective electron-phonon coupling constant $\lambda/J=0.2$ as an equilibrium phase. The equilibrium mean-field parameters are $X_x=X_y=-0.75$ and $\phi=0.27$. (a-c) The collective amplitude mode and (d-f) the collective phase mode of the excitonic phase under BCL, CL and LL, respectively. The dashed gray line marks the light cutoff time, which is chosen at $t=1.0$ ps here.} \label{fig.5}
\end{figure*}

In the following, we will focus on how the BCL affects the behavior of the phase and amplitude of the exciton order parameter. This analysis can also be applied to two other polarizations. When the light is on, the BCL excites electrons from the valence to the conduction band and may increase the exciton density so long as the energy gap remains close to the equilibrium value. The direction and strength of excitation distribution in momentum space depends on the light's polarization and intensity, respectively. From Fig.~\ref{FB}(a,b), at low intensity ($A=1 \mathrm{\mu V.s.m^{-1}}$ and $A=2 \mathrm{\mu V.s.m^{-1}}$) the rotational symmetry breaking is weak, which leads to a small change in the amplitude of the exciton order parameter. In this condition, the coupling term $V\cot(2\gamma_k)\sin(2\gamma_p)$ dominates the correlation between the phases of excitons in different modes. Therefore, the phase of excitons at each mode tends to be aligned and change slightly over the Brillouin zone. This also leads to an enhanced amplitude of the exciton order parameter compared to the equilibrium state, as shown in equation \eqref{eqamp}. At high intensities, there is a threshold intensity (which is $A\simeq 2.2 \mathrm{\mu V.s.m^{-1}}$ for BCL) yielding a drop in the amplitude of the exciton order parameter when the light intensity is increased. On the other hand, as shown in Fig.~\ref{fig.4}, for $A=3 \mathrm{\mu V.s.m^{-1}}$ the symmetry breaking is stronger and the coupling term $V\cot(2\gamma_k)\sin(2\gamma_p)$ varies more than that at $A=2 \mathrm{\mu V.s.m^{-1}}$. Thus, the last term of Eq.\eqref{eq:28} becomes weaker for $A=3 \mathrm{\mu V.s.m^{-1}}$. Therefore, at $A=3\mathrm{\mu V.s.m^{-1}}$ the phase of excitons can change more easily in momentum space, which results in a larger phase difference between excitons and a lower amplitude intensity according to equation \eqref{eqamp}. In short, the light polarization and intensity affect the excitonic phase correlation and hence the rate of change of the phase and the oscillation magnitude of the order parameter.

 Furthermore, in switch OFF region ($A=0$), equation \eqref{eq:28} implies that each exciton in Brillouin zone has an individual frequency $\omega_k=B_k^z$. Amplitude of the exciton order parameter $\phi_{ex}(t)$ also changes according to light intensity before turning off. This is due to memory effect in system, i.e., the current state of the system depends on the previous state. Therefore the amplitude $\phi_{ex}(t)$ decrease less for low light intensity than for high intensity, because of correlations between exciton phases.  Moreover, from the last term of equation \eqref{eq:28}, the evolution of collective modes in switch ON and OFF regions varies with Coulomb interaction strength $V$ and shows different behaviors in BCS and BEC regimes of the  exciton phase diagram.

\section{The effects of electron-phonon coupling} \label{sec:el-ph}
Phonons can modify the dynamics of excitations in excitonic insulators \cite{murakami2020collective, golevz2020nonlinear,khatibi2020excitonic,bretscher2021imaging,baldini2023spontaneous,zenker2014fate,kaneko2013orthorhombic,PWerner}, and hence their indispensable role has to be taken into account. In this section, we investigate the effect of electron - phonon coupling on the exciton condensation by considering an optical branch of phonon modes that modifies the inter-orbital hybridization. The Hamiltonian of the electron - phonon coupling is given by
	
\begin{align}
\hat{H}_{\mathrm{e-ph}}=&g_x\sum_j \big(\hat{b}_{j,x}^\dagger+\hat{b}_{j,x}\big)\big(\hat{c}^\dagger_{j,0}\hat{c}_{j,1}+\hat{c}^\dagger_{j,1}\hat{c}_{j,0}\big)\nonumber\\
	&+g_y\sum_j \big(\hat{b}_{j,y}^\dagger+\hat{b}_{j,y}\big)\big(\hat{c}^\dagger_{j,0}\hat{c}_{j,1}+\hat{c}^\dagger_{j,1}\hat{c}_{j,0}\big),\label{eq:31}
	\end{align} 
	and the phonons are described by 
	\begin{equation}
	\label{Hph}
	\hat{H}_{ph}=\omega_{ph_x}\sum_j \hat{b}_{j,x}^\dagger\hat{b}_{j,x}+\omega_{ph_y}\sum_j \hat{b}_{j,y}^\dagger\hat{b}_{j,y}.
	\end{equation}
	
Here, $\omega_{ph_\delta}$ is the phonon frequency which comes from the vibration of the atoms in the $\delta$-direction ($\delta=x,y$). $\hat{b}_{j,\delta}^\dagger$ is the phonon creation operator at site $j$ and $g_\delta$ is the electron-phonon coupling strength in the $\delta$-direction. The displacement and momentum operators read as $\hat{X}_{j\delta}=\hat{b}_{j,\delta}^\dagger+\hat{b}_{j,\delta}$ and $\hat{P}_{j\delta}=i(\hat{b}_{j,\delta}^\dagger-\hat{b}_{j,\delta})$, respectively. For simplicity, we assume that the frequency and the strength of the electron - phonon coupling in two directions $x$ and $y$ are the same, $\omega_{ph_\delta}=\omega_{ph}$ and $g_\delta=g$, and define the effective electron - phonon coupling as $\lambda\equiv \frac{2g^2}{\omega_{ph}}$. This coupling cooperates with the Coulomb interaction and leads to an exciton phase transition in the system \cite{kaneko2013orthorhombic,golevz2020nonlinear,murakami2020collective,zenker2014fate,PWerner}. 
	
In the absence of electron - phonon coupling, the system described by Hamiltonian \eqref{eq:1} is manifestly $\mathrm{U(1)}\times\mathrm{U(1)}$ symmetric. It is straightforward to see that the Hamiltonian is invariant under separate global phase rotations of electron operators in valence and conduction bands, $\hat{c}_{i,\alpha}\rightarrow e^{i\varphi_{\alpha}}\hat{c}_{i,\alpha}$, hence leading to separate charge conservations in the bands. By changing the phase variables to total $\varphi_{t}=\varphi_0+\varphi_1$ and relative $\varphi_{r}=\varphi_0-\varphi_1$, the global symmetry casts into $\mathrm{U(1)}_{t}\times\mathrm{U(1)}_{r}$. Upon  exciton condensation, i.e., the development of $\phi\neq0$ in the system, the global symmetry breaks down to a single $\mathrm{U(1)}_t$ symmetry of total charge conservation. Consequently, the phase fluctuations of $\phi$ remains gapless akin to the Goldeston modes. Coupling to phonons, according to equation \eqref{eq:31}, breaks the relative $\mathrm{U(1)}_r$ symmetry down to discrete $Z_2$ symmetry at the level of Hamiltonian explicitly, which results in a gapped phase mode in the exciton condensation \cite{yusupov2010coherent,matsunaga2017polarization,golevz2020nonlinear,murakami2020collective, zenker2014fate,sun2020bardasis}. 
	
We study the dynamics of the system in the presence of the phonon and in the pseudo-spin representation for different light polarizations by considering Hamiltonian \eqref{eq:31} together with Hamiltonian \eqref{eq:1}, $\tilde{H}=\hat{H}+ \hat{H}_{\mathrm{e-ph}}$, and treat both $\hat{H}_{int}$ and $\hat{H}_{\mathrm{e-ph}}$ using time-dependent mean-field theory \cite{PWerner,murakami2020collective}. In the presence of the electron-phonon coupling, the $x$ component of pseudo-magnetic field becomes $B^x_k=2\left(g\left( X_x(t)+X_y(t)\right)-V\right)\mathrm{Re}[\phi(t)]$, and $dP_\delta(t)/dt=-\omega_{\mathrm{ph}} X_\delta(t)-4g \Re[\phi(t)]$ and $dX_\delta(t)/dt=\omega_{\mathrm{ph}}P_\delta(t)$. 
	

The dynamics of the amplitude and the phase of exciton order parameter in the presence of phonons for different light polarizations is shown in Fig.~\ref{fig.5}. We set the parameters as $V/J=4$, $D/J=0.9$, $\lambda/J=0.2$ and $\omega_{ph}/J=0.1$ such that the system is prepared in the BCS regime of the exciton phase where the equilibrium mean-field parameters are $X_x=X_y=-0.75$ and $\phi=0.27$
with the energy gap $E_g=2\left(2g(X_x+X_y)-V\right)\Re[\phi(0)]/J=2.34$. Again, we use the Kuramoto model to describe the evolution of the collective modes. The equation \eqref{eq:24} becomes
	
\begin{align}
\dot{\theta}_k=&\left(\epsilon_{k+\frac{e}{\hbar}A(t),0}-\epsilon_{k+\frac{e}{\hbar}A(t),1}+V(n_1-n_0)\right)\nonumber\\
&+\frac{V}{N}\cot(2\gamma_k)\sum_p\sin(2\gamma_p)\cos(\theta_p-\theta_k)\nonumber\\
&+2g\left( X_x(t)+ X_y(t)\right) \cot(2\gamma_k) \cos(\theta_k),\label{eq:35}
\end{align}
and equation \eqref{eq:25} evolves as
\begin{align}
\dot{\gamma}_k=&\frac{-V}{2N}\sum_p\sin(2\gamma_p)\sin(\theta_p-\theta_k)\nonumber\\
&-g\left( X_x(t)+ X_y(t)\right)\sin(\theta_k).\label{eq:36}
\end{align}
	
\begin{figure*}[t]
\includegraphics[width=0.85\textwidth]{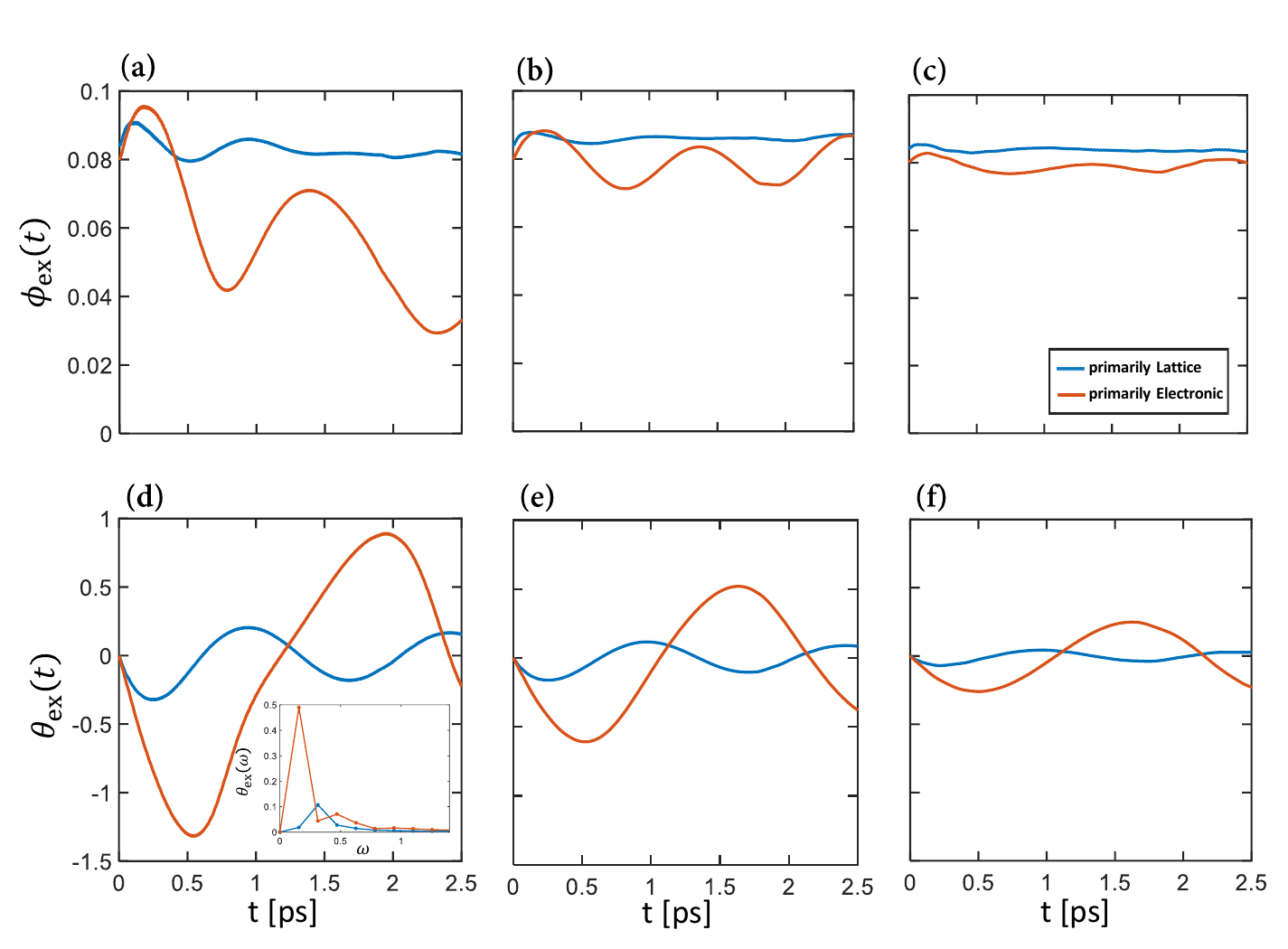}
\caption{The dynamics of the exciton order parameter in two different cases are compared: primarily electronic and primarily lattice. We use a high frequency light field with $\Omega=0.3~ \mathrm{eV}$ and $A=1.0 ~\mathrm{\mu V.s.m^{-1}}$ and different polarizations. The model parameters are $V/J=1.58$, $D/J=0.9$, $\lambda/J=0.08$ for the primarily electronic case, and $V/J=0.6$, $D/J=0.9$, $\lambda/J=0.3$ for the primarily lattice case. The phonon frequency is $\omega_{\mathrm{ph}}/J=0.1$ in both cases. The equilibrium mean field parameters are  $\phi=0.081$ and $X_x=X_y=-0.2$ for the primarily electronic case, and $\phi=0.085$  and $X_x=X_y=-0.41$ for the primarily lattice case, which fix the equilibrium gap $E_g/J=0.3$ in both cases.  Figures (a-c) show the collective amplitude mode and figures (d-f) show the collective phase mode of the excitonic phase under BCL, CL and LL, respectively. The inset of Fig.d displays the Fourier spectrum of the exciton phase under a BCL, where the oscillation frequency is $\omega=0.31 ~\mathrm{eV}$ for the primarily lattice case and $\omega=0.16~ \mathrm{eV}$ for the primarily electronic case. These figures only show the evolution of the exciton order when the light field is switched ON, without switching it OFF.} \label{fig.7}
\end{figure*}
	
By comparing the Kuramoto model \eqref{eq:30} with equation \eqref{eq:35} we see that the phonons in the system act as an external force with strength $2g\left( X_x(t)+ X_y(t)\right) \cot(2\gamma_k)$, which can affect the synchronization of the phases of excitons at different momenta in various ways, depending on its strength and frequency. 
As we shall discuss below this would affect the dynamics of the phase of the exciton.    
Equation \eqref{eq:35} shows that the synchronization of the phases of excitons in momentum space is affected by the electron-phonon coupling. Let us take BCL polarization as a specific case. Fig.~\ref{fig.5} (a,d) shows the dynamics of the amplitude and phase of the exciton condensation for different light intensities. It is seen that the electron-phonon coupling as an external force induces strong anharmonicity in the amplitude oscillations $\phi_{ex}(t)$ as compared to Fig.~\ref{fig.4}.

As shown in Fig.~\ref{fig.5}(d), the phase of exciton oscillates for $A=1 \mathrm{\mu V.s.m^{-1}}$ but it varies linearly (mode $2\pi$) with time when the light intensity increases to $A=2 \mathrm{\mu V.s.m^{-1}}$. The BCL polarization causes stronger rotational symmetry breaking as the light intensity increases. The last term of Eq.\eqref{eq:35} changes rapidly and most likely does not contribute to the time evolution of the exciton phase, so the phase of order parameter changes linearly with time at high intensities similar to the case without electron-phonon coupling. At lower intensity $A=1 \mathrm{\mu V.s.m^{-1}}$, the phase evolution follows an oscillatory behavior, associated with the U(1) symmetry breaking in \eqref{eq:31} as discussed before. Upon symmetry breaking, the phase mode acquires a mass and its dynamics is governed by a harmonic oscillator equation $\ddot{\theta}_{ex}+\omega_{\theta}^{2}\theta_{ex}=0$ derived from low-enegry effective theory in Ref.[\onlinecite{golevz2020nonlinear}]. From Fig.~\ref{fig.5}(e-f), we see that the phase response also depends on the light polarization. For instance, for the linear polarization the phase oscillation is observed even for intensive fields.

\section{Evolution of the collective modes in electronic and lattice driven excitonic condensates} \label{sec:PLPE}
The weak and strong electron-phonon coupling considerably modify the nonlinear optical response of the excitonic system \cite{golevz2020nonlinear} in the pump-probe measurements by exciting the phase and phonon modes, respectively. In the primarily electronic driven case, where the U(1) symmetry is broken weakly due to weak electron-phonon interaction leading to a small gap of the phase mode, the pump pulse stimulates the low-energy phase modes which reflected as a small peak in the nonlinear optical conductivity. For the case of strong U(1) symmetry breaking, occurring when the electron-phonon interaction is strong, the low-energy spectra is accumulated by the phononic excitations and the phase modes appear at much higher energies. 
	
Here, we examine the dynamics of the exciton order parameter in primarily electronic and primarily lattice cases, for which the initial parameters were chosen in the BCS regime such that the single particle gap ($E_{g}/J=0.3$) of the system is the same for both cases. Fig.~\ref{fig.7} shows the time evolution of amplitude and phase modes. We set the parameters as $V/J=1.58$, $D/J=0.9$, $\lambda/J=0.08$, $\omega_{\mathrm{ph}}/J=0.1$ for the primarily electronic and $V/J=0.6$, $D/J=0.9$, $\lambda/J=0.3$, $\omega_{\mathrm{ph}}/J=0.1$ for the primarily lattice case with $A=1 \mathrm{\mu V.s.m^{-1}}$.
	
As shown in Fig.~\ref{fig.7}(a-c), the amplitude of the exciton order parameter does not evolve considerably for all types of polarization. Instead, as shown in Fig.~\ref{fig.7}(d-f) the phase mode $\theta_{\mathrm{ex}}(t)$ evolves almost harmonically described by $\ddot{\theta}_{ex}+\omega_{\theta}^{2}\theta_{\mathrm{ex}}=0$ \cite{golevz2020nonlinear} as we mentioned in the preceding section. By Fourier transforming the data in Fig.~\ref{fig.7}(d-f) (see the inset), we extract the oscillation frequency and the gap of the phase mode. It reads as $\omega=0.31$ and $\omega=0.16$ for the primarily lattice and electronic cases, respectively. These results also show that the amplitude of the phase oscillations for BCL and CL is larger than that of the LL. Thus, the phase mode signals in the pump-probe measurements could be more pronounced for former polarizations, providing a way to distinguish between the microscopic origin of the exciton formation, either purely electronic or electron-lattice interactions \cite{murakami2020collective} in candidate materials. 
	
\section{conclusions \label{conclusion}}
This work is mainly motivated by the discovery of possible excitonic insulator phase in layered dichalcogenide Ta$_2$NiSe$_5$, attracted considerable attention both experimentally and theoretically in recent years. Besides the structural phase transition, the microscopic origin of the formation of excitons of being mediated by purely electronic correlations or by phonons is still controversial. Therefore, it's highly demanding to explore the signature of the excitonic condensate particularly in optical responses. 
	
Here, we studied dynamical properties of the condensate when the system is periodically driven in time. We considered periodic drives with light polarizations such as bicircular, circular, and linear, and different intensities. In the absence of coupling to phonons, we found that the amplitude mode of the condensate oscillates in time with a frequency set by the insulating gap. Assuming the modes are weakly dissipated, the modes keep oscillating almost harmonically even after the light is switched off. The gapless phase mode evolves linearly in time, acting as a rotor moving around a circle with a constant angular velocity depending on the polarization and intensity of the drive. 
	
Coupling the electronic bands to phonons gaps out the phase mode due to symmetry breaking. We found that, while for intensive optical fields it may still be evolved linearly in time for bicircular polarization, in most cases the time evolution of the phase mode follows a harmonic oscillator. In fact, the electron-phonon coupling provides a trapping potential for the phase mode leading to a harmonic oscillator behavior. Furthermore, we establish that the weak versus strong electron-phonon coupling has considerable effects on the time evolution of the condensate. For the former case, where the electronic correlations are dominant, the dynamics of the amplitude and phase modes are more susceptible to the polarization of the drive. We speculate that this time evolution can have profound effects on the optical response and induce features in the reflectivity measurments of the excitonic insulators, and may serve as a possible probe to get insight into the ground state of material candidates.          
	
\section{Acknowledgement}
The authors would like to thank Sharif University of Technology for supports.

\appendix 
\section{Floquet theory} \label{APP_A}
Floquet theory is a suitable approach for solving a time-periodic Hamiltonian $\hat{H}(t)=\hat{H}(t+T)$, where $T$ is a period of the drive and is related to the drive frequency as $\Omega=\frac{2\pi}{T}$ \cite{FloquetSolution}. According to the Floquet formalism, the solution of the time-dependent Schr\"{o}dinger equation 
$i\hbar\frac{d}{dt}|\Psi_\alpha(t)\rangle=\hat{H}(t)|\Psi_\alpha(t)\rangle$ with a time-periodic Hamiltonian is a complete set of orthogonal solutions which can be written as the product of a plane wave and a periodic function of the form
$|\Psi_\alpha(t)\rangle=\exp(-i\varepsilon_\alpha t/\hbar)|\Phi_\alpha(t)\rangle$ \cite{autler1955stark,FloquetSolution,zel1967quasienergy,sambe1973steady}. 
The periodic function $|\Phi_\alpha(t)\rangle$
is a Floquet state with a period of the Hamiltonian and satisfies the Floquet-Schr\"{o}dinger equation as follows
	\begin{align}
	H_F(t)|\Phi_\alpha(t)\rangle=\varepsilon_\alpha |\Phi_\alpha(t)\rangle, \label{eq:A1}
	\end{align}
where the Floquet Hamiltonian is defined by 
	\begin{align}
	H_F(t)\equiv\hat{H}(t)-i\hbar\frac{d}{dt}.\label{eq:A2}
	\end{align}

	From the above equations, 
	$|\Phi_\alpha(t)\rangle$
	can also be considered as an eigenstate of the time-dependent Floquet Hamiltonian with a time-independent eigenvalue
	$\varepsilon_\alpha$. Here  
	$\varepsilon_\alpha$ is a quasienergy of the system and, compared to the first Brillouin zone in Bloch theory for a particle moving under a periodic potential in a real space, all solutions of the Floquet-Schr\"{o}dinger equation are indexed by these quasienergies belonging to the first Floquet zone
	$\varepsilon_\alpha\in[-\Omega/2,\Omega/2]$ \cite{Floquetreview,FloquetSolution,sambe1973steady}.
	
Solving Eq.\eqref{eq:A1} is often rather challenging, but due to the periodicity of  Floquet state and Floquet Hamiltonian, we can expand them in a Fourier series. Doing so, the Eq.\eqref{eq:A1} is formulated as an infinite dimensional eigenvalue problem
\begin{align}
\sum_{m=-\infty}^{+\infty}\biggl(\hat{H}_{n-m}-n\hbar\Omega\delta_{mn}\biggr)|\Phi^n_\alpha\rangle=\varepsilon_\alpha |\Phi_\alpha^n\rangle\label{eq:A3}
\end{align}
where
$\hat{H}(t)=\sum_{n=-\infty}^{\infty}e^{-in\Omega t}\hat{H}_n$ and $|\Phi_\alpha(t)\rangle=\sum_{n=-\infty}^{\infty}e^{-in\Omega t}|\Phi^n_\alpha\rangle$
are the Fourier transform of the Hamiltonian and Floquet states, respectively. For numerical calculations the effective Hamiltonian $(H_{eff})_{nm}=H_{n-m}-n\hbar\Omega\delta_{mn}$ is truncated by some approximations and perturbation theory \cite{FloquetGoldman,Floquetreview,rahav2003effective,casas2001floquet,blanes2009magnus,mananga2011introduction,mikami2016brillouin}.

\section{Comment on errors in RK method} \label{B}
The time dependent variables $\phi(t)$, $n_\alpha(t)$ ($\alpha=0,1$) and $X_\delta(t) $ ($\delta=x,y$) were calculated by numerically solving the Heisenberg equations of motion (Eq.\eqref{eq:20}, main text) using the fourth-order Runge-Kutta  method. Here below, we explain more about the error analysis that was done for the fourth-order RK method. We purposely applied the step doubling technique, as the most straightforward technique for the adaptive step size control ~\cite{press1992adaptive}, to get a good accuracy in the solution. The local truncation error, i.e. the error induced for each successive stage of the iterated algorithm, was calculated twice: a full step, with a step size of $\delta t=0.0001 \mathrm{ps}$, then, independently, two half-steps.
The deviation of the local truncation error for different light polarizations (LL, CL, BCL) is less than $1\%$.


\begin{thebibliography}{72}
		\expandafter\ifx\csname natexlab\endcsname\relax\def\natexlab#1{#1}\fi
		\expandafter\ifx\csname bibnamefont\endcsname\relax
		\def\bibnamefont#1{#1}\fi
		\expandafter\ifx\csname bibfnamefont\endcsname\relax
		\def\bibfnamefont#1{#1}\fi
		\expandafter\ifx\csname citenamefont\endcsname\relax
		\def\citenamefont#1{#1}\fi
		\expandafter\ifx\csname url\endcsname\relax
		\def\url#1{\texttt{#1}}\fi
		\expandafter\ifx\csname urlprefix\endcsname\relax\def\urlprefix{URL }\fi
		\providecommand{\bibinfo}[2]{#2}
		\providecommand{\eprint}[2][]{\url{#2}}
		
		\bibitem[{\citenamefont{J{\'e}rome et~al.}(1967)\citenamefont{J{\'e}rome, Rice,
				and Kohn}}]{jerome1967excitonic}
		\bibinfo{author}{\bibfnamefont{D.}~\bibnamefont{J{\'e}rome}},
		\bibinfo{author}{\bibfnamefont{T.}~\bibnamefont{Rice}}, \bibnamefont{and}
		\bibinfo{author}{\bibfnamefont{W.}~\bibnamefont{Kohn}},
		\bibinfo{journal}{Physical Review} \textbf{\bibinfo{volume}{158}},
		\bibinfo{pages}{462} (\bibinfo{year}{1967}).
		
		\bibitem[{\citenamefont{Halperin and Rice}(1968)}]{halperin1968possible}
		\bibinfo{author}{\bibfnamefont{B.}~\bibnamefont{Halperin}} \bibnamefont{and}
		\bibinfo{author}{\bibfnamefont{T.}~\bibnamefont{Rice}},
		\bibinfo{journal}{Reviews of Modern Physics} \textbf{\bibinfo{volume}{40}},
		\bibinfo{pages}{755} (\bibinfo{year}{1968}).
		
		\bibitem[{\citenamefont{Kohn}(1967)}]{kohn1967excitonic}
		\bibinfo{author}{\bibfnamefont{W.}~\bibnamefont{Kohn}},
		\bibinfo{journal}{Physical Review Letters} \textbf{\bibinfo{volume}{19}},
		\bibinfo{pages}{439} (\bibinfo{year}{1967}).
		
		\bibitem[{\citenamefont{Keldysh and Kozlov}(1968)}]{keldysh1968collective}
		\bibinfo{author}{\bibfnamefont{L.}~\bibnamefont{Keldysh}} \bibnamefont{and}
		\bibinfo{author}{\bibfnamefont{A.}~\bibnamefont{Kozlov}},
		\bibinfo{journal}{Sov. Phys. JETP} \textbf{\bibinfo{volume}{27}},
		\bibinfo{pages}{521} (\bibinfo{year}{1968}).
		
		\bibitem[{\citenamefont{Combescot and Shiau}(2015)}]{combescot2015excitons}
		\bibinfo{author}{\bibfnamefont{M.}~\bibnamefont{Combescot}} \bibnamefont{and}
		\bibinfo{author}{\bibfnamefont{S.-Y.} \bibnamefont{Shiau}},
		\emph{\bibinfo{title}{Excitons and Cooper pairs: two composite bosons in
				many-body physics}} (\bibinfo{publisher}{Oxford University Press},
		\bibinfo{year}{2015}).
		
		\bibitem[{\citenamefont{Ma et~al.}(2021)\citenamefont{Ma, Nguyen, Wang, Zeng,
				Watanabe, Taniguchi, MacDonald, Mak, and Shan}}]{ma2021strongly}
		\bibinfo{author}{\bibfnamefont{L.}~\bibnamefont{Ma}},
		\bibinfo{author}{\bibfnamefont{P.~X.} \bibnamefont{Nguyen}},
		\bibinfo{author}{\bibfnamefont{Z.}~\bibnamefont{Wang}},
		\bibinfo{author}{\bibfnamefont{Y.}~\bibnamefont{Zeng}},
		\bibinfo{author}{\bibfnamefont{K.}~\bibnamefont{Watanabe}},
		\bibinfo{author}{\bibfnamefont{T.}~\bibnamefont{Taniguchi}},
		\bibinfo{author}{\bibfnamefont{A.~H.} \bibnamefont{MacDonald}},
		\bibinfo{author}{\bibfnamefont{K.~F.} \bibnamefont{Mak}}, \bibnamefont{and}
		\bibinfo{author}{\bibfnamefont{J.}~\bibnamefont{Shan}},
		\bibinfo{journal}{Nature} \textbf{\bibinfo{volume}{598}},
		\bibinfo{pages}{585} (\bibinfo{year}{2021}).
		
		\bibitem[{\citenamefont{Ataei et~al.}(2021)\citenamefont{Ataei, Varsano,
				Molinari, and Rontani}}]{ataei2021evidence}
		\bibinfo{author}{\bibfnamefont{S.~S.} \bibnamefont{Ataei}},
		\bibinfo{author}{\bibfnamefont{D.}~\bibnamefont{Varsano}},
		\bibinfo{author}{\bibfnamefont{E.}~\bibnamefont{Molinari}}, \bibnamefont{and}
		\bibinfo{author}{\bibfnamefont{M.}~\bibnamefont{Rontani}},
		\bibinfo{journal}{Proceedings of the National Academy of Sciences}
		\textbf{\bibinfo{volume}{118}}, \bibinfo{pages}{e2010110118}
		(\bibinfo{year}{2021}).
		
		\bibitem[{\citenamefont{Sun et~al.}(2022)\citenamefont{Sun, Zhao, Palomaki,
				Fei, Runburg, Malinowski, Huang, Cenker, Cui, Chu et~al.}}]{sun2022evidence}
		\bibinfo{author}{\bibfnamefont{B.}~\bibnamefont{Sun}},
		\bibinfo{author}{\bibfnamefont{W.}~\bibnamefont{Zhao}},
		\bibinfo{author}{\bibfnamefont{T.}~\bibnamefont{Palomaki}},
		\bibinfo{author}{\bibfnamefont{Z.}~\bibnamefont{Fei}},
		\bibinfo{author}{\bibfnamefont{E.}~\bibnamefont{Runburg}},
		\bibinfo{author}{\bibfnamefont{P.}~\bibnamefont{Malinowski}},
		\bibinfo{author}{\bibfnamefont{X.}~\bibnamefont{Huang}},
		\bibinfo{author}{\bibfnamefont{J.}~\bibnamefont{Cenker}},
		\bibinfo{author}{\bibfnamefont{Y.-T.} \bibnamefont{Cui}},
		\bibinfo{author}{\bibfnamefont{J.-H.} \bibnamefont{Chu}},
		\bibnamefont{et~al.}, \bibinfo{journal}{Nature Physics}
		\textbf{\bibinfo{volume}{18}}, \bibinfo{pages}{94} (\bibinfo{year}{2022}).
		
		\bibitem[{\citenamefont{Varsano et~al.}(2020)\citenamefont{Varsano, Palummo,
				Molinari, and Rontani}}]{varsano2020monolayer}
		\bibinfo{author}{\bibfnamefont{D.}~\bibnamefont{Varsano}},
		\bibinfo{author}{\bibfnamefont{M.}~\bibnamefont{Palummo}},
		\bibinfo{author}{\bibfnamefont{E.}~\bibnamefont{Molinari}}, \bibnamefont{and}
		\bibinfo{author}{\bibfnamefont{M.}~\bibnamefont{Rontani}},
		\bibinfo{journal}{Nature nanotechnology} \textbf{\bibinfo{volume}{15}},
		\bibinfo{pages}{367} (\bibinfo{year}{2020}).
		
		\bibitem[{\citenamefont{Kogar et~al.}(2017)\citenamefont{Kogar, Rak, Vig,
				Husain, Flicker, Joe, Venema, MacDougall, Chiang, Fradkin
				et~al.}}]{kogar2017signatures}
		\bibinfo{author}{\bibfnamefont{A.}~\bibnamefont{Kogar}},
		\bibinfo{author}{\bibfnamefont{M.~S.} \bibnamefont{Rak}},
		\bibinfo{author}{\bibfnamefont{S.}~\bibnamefont{Vig}},
		\bibinfo{author}{\bibfnamefont{A.~A.} \bibnamefont{Husain}},
		\bibinfo{author}{\bibfnamefont{F.}~\bibnamefont{Flicker}},
		\bibinfo{author}{\bibfnamefont{Y.~I.} \bibnamefont{Joe}},
		\bibinfo{author}{\bibfnamefont{L.}~\bibnamefont{Venema}},
		\bibinfo{author}{\bibfnamefont{G.~J.} \bibnamefont{MacDougall}},
		\bibinfo{author}{\bibfnamefont{T.~C.} \bibnamefont{Chiang}},
		\bibinfo{author}{\bibfnamefont{E.}~\bibnamefont{Fradkin}},
		\bibnamefont{et~al.}, \bibinfo{journal}{Science}
		\textbf{\bibinfo{volume}{358}}, \bibinfo{pages}{1314} (\bibinfo{year}{2017}).
		
		\bibitem[{\citenamefont{Chernikov et~al.}(2014)\citenamefont{Chernikov,
				Berkelbach, Hill, Rigosi, Li, Aslan, Reichman, Hybertsen, and
				Heinz}}]{chernikov2014excitons}
		\bibinfo{author}{\bibfnamefont{A.}~\bibnamefont{Chernikov}},
		\bibinfo{author}{\bibfnamefont{T.~C.} \bibnamefont{Berkelbach}},
		\bibinfo{author}{\bibfnamefont{H.~M.} \bibnamefont{Hill}},
		\bibinfo{author}{\bibfnamefont{A.}~\bibnamefont{Rigosi}},
		\bibinfo{author}{\bibfnamefont{Y.}~\bibnamefont{Li}},
		\bibinfo{author}{\bibfnamefont{{\"O}.~B.} \bibnamefont{Aslan}},
		\bibinfo{author}{\bibfnamefont{D.~R.} \bibnamefont{Reichman}},
		\bibinfo{author}{\bibfnamefont{M.~S.} \bibnamefont{Hybertsen}},
		\bibnamefont{and} \bibinfo{author}{\bibfnamefont{T.~F.} \bibnamefont{Heinz}},
		in \emph{\bibinfo{booktitle}{2014 Conference on Lasers and Electro-Optics
				(CLEO)-Laser Science to Photonic Applications}}
		(\bibinfo{organization}{IEEE}, \bibinfo{year}{2014}), pp.
		\bibinfo{pages}{1--2}.
		
		\bibitem[{\citenamefont{Behura et~al.}(2021)\citenamefont{Behura, Miranda,
				Nayak, Johnson, Das, and Pradhan}}]{behura2021moire}
		\bibinfo{author}{\bibfnamefont{S.~K.} \bibnamefont{Behura}},
		\bibinfo{author}{\bibfnamefont{A.}~\bibnamefont{Miranda}},
		\bibinfo{author}{\bibfnamefont{S.}~\bibnamefont{Nayak}},
		\bibinfo{author}{\bibfnamefont{K.}~\bibnamefont{Johnson}},
		\bibinfo{author}{\bibfnamefont{P.}~\bibnamefont{Das}}, \bibnamefont{and}
		\bibinfo{author}{\bibfnamefont{N.~R.} \bibnamefont{Pradhan}},
		\bibinfo{journal}{Emergent Materials} \textbf{\bibinfo{volume}{4}},
		\bibinfo{pages}{813} (\bibinfo{year}{2021}).
		
		\bibitem[{\citenamefont{Wang et~al.}(2018)\citenamefont{Wang, Chernikov,
				Glazov, Heinz, Marie, Amand, and Urbaszek}}]{wang2018colloquium}
		\bibinfo{author}{\bibfnamefont{G.}~\bibnamefont{Wang}},
		\bibinfo{author}{\bibfnamefont{A.}~\bibnamefont{Chernikov}},
		\bibinfo{author}{\bibfnamefont{M.~M.} \bibnamefont{Glazov}},
		\bibinfo{author}{\bibfnamefont{T.~F.} \bibnamefont{Heinz}},
		\bibinfo{author}{\bibfnamefont{X.}~\bibnamefont{Marie}},
		\bibinfo{author}{\bibfnamefont{T.}~\bibnamefont{Amand}}, \bibnamefont{and}
		\bibinfo{author}{\bibfnamefont{B.}~\bibnamefont{Urbaszek}},
		\bibinfo{journal}{Reviews of Modern Physics} \textbf{\bibinfo{volume}{90}},
		\bibinfo{pages}{021001} (\bibinfo{year}{2018}).
		
		\bibitem[{\citenamefont{Mueller and Malic}(2018)}]{mueller2018exciton}
		\bibinfo{author}{\bibfnamefont{T.}~\bibnamefont{Mueller}} \bibnamefont{and}
		\bibinfo{author}{\bibfnamefont{E.}~\bibnamefont{Malic}},
		\bibinfo{journal}{npj 2D Materials and Applications}
		\textbf{\bibinfo{volume}{2}}, \bibinfo{pages}{29} (\bibinfo{year}{2018}).
		
		\bibitem[{\citenamefont{Di~Salvo et~al.}(1986)\citenamefont{Di~Salvo, Chen,
				Fleming, Waszczak, Dunn, Sunshine, and Ibers}}]{di1986physical}
		\bibinfo{author}{\bibfnamefont{F.}~\bibnamefont{Di~Salvo}},
		\bibinfo{author}{\bibfnamefont{C.}~\bibnamefont{Chen}},
		\bibinfo{author}{\bibfnamefont{R.}~\bibnamefont{Fleming}},
		\bibinfo{author}{\bibfnamefont{J.}~\bibnamefont{Waszczak}},
		\bibinfo{author}{\bibfnamefont{R.}~\bibnamefont{Dunn}},
		\bibinfo{author}{\bibfnamefont{S.}~\bibnamefont{Sunshine}}, \bibnamefont{and}
		\bibinfo{author}{\bibfnamefont{J.~A.} \bibnamefont{Ibers}},
		\bibinfo{journal}{Journal of the Less Common Metals}
		\textbf{\bibinfo{volume}{116}}, \bibinfo{pages}{51} (\bibinfo{year}{1986}).
		
		\bibitem[{\citenamefont{Wakisaka et~al.}(2009)\citenamefont{Wakisaka, Sudayama,
				Takubo, Mizokawa, Arita, Namatame, Taniguchi, Katayama, Nohara, and
				Takagi}}]{wakisaka2009excitonic}
		\bibinfo{author}{\bibfnamefont{Y.}~\bibnamefont{Wakisaka}},
		\bibinfo{author}{\bibfnamefont{T.}~\bibnamefont{Sudayama}},
		\bibinfo{author}{\bibfnamefont{K.}~\bibnamefont{Takubo}},
		\bibinfo{author}{\bibfnamefont{T.}~\bibnamefont{Mizokawa}},
		\bibinfo{author}{\bibfnamefont{M.}~\bibnamefont{Arita}},
		\bibinfo{author}{\bibfnamefont{H.}~\bibnamefont{Namatame}},
		\bibinfo{author}{\bibfnamefont{M.}~\bibnamefont{Taniguchi}},
		\bibinfo{author}{\bibfnamefont{N.}~\bibnamefont{Katayama}},
		\bibinfo{author}{\bibfnamefont{M.}~\bibnamefont{Nohara}}, \bibnamefont{and}
		\bibinfo{author}{\bibfnamefont{H.}~\bibnamefont{Takagi}},
		\bibinfo{journal}{Physical review letters} \textbf{\bibinfo{volume}{103}},
		\bibinfo{pages}{026402} (\bibinfo{year}{2009}).
		
		\bibitem[{\citenamefont{Kaneko et~al.}(2012)\citenamefont{Kaneko, Seki, and
				Ohta}}]{kaneko2012excitonic}
		\bibinfo{author}{\bibfnamefont{T.}~\bibnamefont{Kaneko}},
		\bibinfo{author}{\bibfnamefont{K.}~\bibnamefont{Seki}}, \bibnamefont{and}
		\bibinfo{author}{\bibfnamefont{Y.}~\bibnamefont{Ohta}},
		\bibinfo{journal}{Physical Review B} \textbf{\bibinfo{volume}{85}},
		\bibinfo{pages}{165135} (\bibinfo{year}{2012}).
		
		\bibitem[{\citenamefont{Seki et~al.}(2014)\citenamefont{Seki, Wakisaka, Kaneko,
				Toriyama, Konishi, Sudayama, Saini, Arita, Namatame, Taniguchi
				et~al.}}]{seki2014excitonic}
		\bibinfo{author}{\bibfnamefont{K.}~\bibnamefont{Seki}},
		\bibinfo{author}{\bibfnamefont{Y.}~\bibnamefont{Wakisaka}},
		\bibinfo{author}{\bibfnamefont{T.}~\bibnamefont{Kaneko}},
		\bibinfo{author}{\bibfnamefont{T.}~\bibnamefont{Toriyama}},
		\bibinfo{author}{\bibfnamefont{T.}~\bibnamefont{Konishi}},
		\bibinfo{author}{\bibfnamefont{T.}~\bibnamefont{Sudayama}},
		\bibinfo{author}{\bibfnamefont{N.}~\bibnamefont{Saini}},
		\bibinfo{author}{\bibfnamefont{M.}~\bibnamefont{Arita}},
		\bibinfo{author}{\bibfnamefont{H.}~\bibnamefont{Namatame}},
		\bibinfo{author}{\bibfnamefont{M.}~\bibnamefont{Taniguchi}},
		\bibnamefont{et~al.}, \bibinfo{journal}{Physical Review B}
		\textbf{\bibinfo{volume}{90}}, \bibinfo{pages}{155116}
		(\bibinfo{year}{2014}).
		
		\bibitem[{\citenamefont{Lu et~al.}(2017)\citenamefont{Lu, Kono, Larkin, Rost,
				Takayama, Boris, Keimer, and Takagi}}]{lu2017zero}
		\bibinfo{author}{\bibfnamefont{Y.}~\bibnamefont{Lu}},
		\bibinfo{author}{\bibfnamefont{H.}~\bibnamefont{Kono}},
		\bibinfo{author}{\bibfnamefont{T.}~\bibnamefont{Larkin}},
		\bibinfo{author}{\bibfnamefont{A.}~\bibnamefont{Rost}},
		\bibinfo{author}{\bibfnamefont{T.}~\bibnamefont{Takayama}},
		\bibinfo{author}{\bibfnamefont{A.}~\bibnamefont{Boris}},
		\bibinfo{author}{\bibfnamefont{B.}~\bibnamefont{Keimer}}, \bibnamefont{and}
		\bibinfo{author}{\bibfnamefont{H.}~\bibnamefont{Takagi}},
		\bibinfo{journal}{Nature communications} \textbf{\bibinfo{volume}{8}},
		\bibinfo{pages}{14408} (\bibinfo{year}{2017}).
		
		\bibitem[{\citenamefont{Jiang et~al.}(2020)\citenamefont{Jiang, Lou, Liu, Li,
				Song, Chang, Duan, and Zhang}}]{jiang2020spin}
		\bibinfo{author}{\bibfnamefont{Z.}~\bibnamefont{Jiang}},
		\bibinfo{author}{\bibfnamefont{W.}~\bibnamefont{Lou}},
		\bibinfo{author}{\bibfnamefont{Y.}~\bibnamefont{Liu}},
		\bibinfo{author}{\bibfnamefont{Y.}~\bibnamefont{Li}},
		\bibinfo{author}{\bibfnamefont{H.}~\bibnamefont{Song}},
		\bibinfo{author}{\bibfnamefont{K.}~\bibnamefont{Chang}},
		\bibinfo{author}{\bibfnamefont{W.}~\bibnamefont{Duan}}, \bibnamefont{and}
		\bibinfo{author}{\bibfnamefont{S.}~\bibnamefont{Zhang}},
		\bibinfo{journal}{Physical Review Letters} \textbf{\bibinfo{volume}{124}},
		\bibinfo{pages}{166401} (\bibinfo{year}{2020}).
		
		\bibitem[{\citenamefont{Matsuzaki et~al.}(2017)\citenamefont{Matsuzaki, Soma,
				Fukuoka, Kodama, Asahara, Suemoto, Adachi, and Uchino}}]{matsuzaki2017purely}
		\bibinfo{author}{\bibfnamefont{R.}~\bibnamefont{Matsuzaki}},
		\bibinfo{author}{\bibfnamefont{H.}~\bibnamefont{Soma}},
		\bibinfo{author}{\bibfnamefont{K.}~\bibnamefont{Fukuoka}},
		\bibinfo{author}{\bibfnamefont{K.}~\bibnamefont{Kodama}},
		\bibinfo{author}{\bibfnamefont{A.}~\bibnamefont{Asahara}},
		\bibinfo{author}{\bibfnamefont{T.}~\bibnamefont{Suemoto}},
		\bibinfo{author}{\bibfnamefont{Y.}~\bibnamefont{Adachi}}, \bibnamefont{and}
		\bibinfo{author}{\bibfnamefont{T.}~\bibnamefont{Uchino}},
		\bibinfo{journal}{Physical Review B} \textbf{\bibinfo{volume}{96}},
		\bibinfo{pages}{125306} (\bibinfo{year}{2017}).
		
		\bibitem[{\citenamefont{Lopes et~al.}(2022)\citenamefont{Lopes, Continentino,
				and Barci}}]{lopes2022excitonic}
		\bibinfo{author}{\bibfnamefont{N.}~\bibnamefont{Lopes}},
		\bibinfo{author}{\bibfnamefont{M.~A.} \bibnamefont{Continentino}},
		\bibnamefont{and} \bibinfo{author}{\bibfnamefont{D.~G.} \bibnamefont{Barci}},
		\bibinfo{journal}{Physical Review B} \textbf{\bibinfo{volume}{105}},
		\bibinfo{pages}{165125} (\bibinfo{year}{2022}).
		
		\bibitem[{\citenamefont{Jiang et~al.}(2019)\citenamefont{Jiang, Li, Duan, and
				Zhang}}]{jiang2019half}
		\bibinfo{author}{\bibfnamefont{Z.}~\bibnamefont{Jiang}},
		\bibinfo{author}{\bibfnamefont{Y.}~\bibnamefont{Li}},
		\bibinfo{author}{\bibfnamefont{W.}~\bibnamefont{Duan}}, \bibnamefont{and}
		\bibinfo{author}{\bibfnamefont{S.}~\bibnamefont{Zhang}},
		\bibinfo{journal}{Physical Review Letters} \textbf{\bibinfo{volume}{122}},
		\bibinfo{pages}{236402} (\bibinfo{year}{2019}).
		
		\bibitem[{\citenamefont{Mak and Shan}(2018)}]{mak2018opportunities}
		\bibinfo{author}{\bibfnamefont{K.~F.} \bibnamefont{Mak}} \bibnamefont{and}
		\bibinfo{author}{\bibfnamefont{J.}~\bibnamefont{Shan}},
		\bibinfo{journal}{Nature nanotechnology} \textbf{\bibinfo{volume}{13}},
		\bibinfo{pages}{974} (\bibinfo{year}{2018}).
		
		\bibitem[{\citenamefont{Combescot et~al.}(2017)\citenamefont{Combescot,
				Combescot, and Dubin}}]{combescot2017bose}
		\bibinfo{author}{\bibfnamefont{M.}~\bibnamefont{Combescot}},
		\bibinfo{author}{\bibfnamefont{R.}~\bibnamefont{Combescot}},
		\bibnamefont{and} \bibinfo{author}{\bibfnamefont{F.}~\bibnamefont{Dubin}},
		\bibinfo{journal}{Reports on Progress in Physics}
		\textbf{\bibinfo{volume}{80}}, \bibinfo{pages}{066501}
		(\bibinfo{year}{2017}).
		
		\bibitem[{\citenamefont{Sun et~al.}(2017)\citenamefont{Sun, Wen, Yoon, Liu,
				Steger, Pfeiffer, West, Snoke, and Nelson}}]{sun2017bose}
		\bibinfo{author}{\bibfnamefont{Y.}~\bibnamefont{Sun}},
		\bibinfo{author}{\bibfnamefont{P.}~\bibnamefont{Wen}},
		\bibinfo{author}{\bibfnamefont{Y.}~\bibnamefont{Yoon}},
		\bibinfo{author}{\bibfnamefont{G.}~\bibnamefont{Liu}},
		\bibinfo{author}{\bibfnamefont{M.}~\bibnamefont{Steger}},
		\bibinfo{author}{\bibfnamefont{L.~N.} \bibnamefont{Pfeiffer}},
		\bibinfo{author}{\bibfnamefont{K.}~\bibnamefont{West}},
		\bibinfo{author}{\bibfnamefont{D.~W.} \bibnamefont{Snoke}}, \bibnamefont{and}
		\bibinfo{author}{\bibfnamefont{K.~A.} \bibnamefont{Nelson}},
		\bibinfo{journal}{Physical review letters} \textbf{\bibinfo{volume}{118}},
		\bibinfo{pages}{016602} (\bibinfo{year}{2017}).
		
		\bibitem[{\citenamefont{Ball}(2022)}]{ball2022ultrafast}
		\bibinfo{author}{\bibfnamefont{P.}~\bibnamefont{Ball}},
		\bibinfo{journal}{Physics} \textbf{\bibinfo{volume}{15}},
		\bibinfo{pages}{118} (\bibinfo{year}{2022}).
		
		\bibitem[{\citenamefont{Bretscher
				et~al.}(2021{\natexlab{a}})\citenamefont{Bretscher, Andrich, Telang, Singh,
				Harnagea, Sood, and Rao}}]{bretscher2021ultrafast}
		\bibinfo{author}{\bibfnamefont{H.~M.} \bibnamefont{Bretscher}},
		\bibinfo{author}{\bibfnamefont{P.}~\bibnamefont{Andrich}},
		\bibinfo{author}{\bibfnamefont{P.}~\bibnamefont{Telang}},
		\bibinfo{author}{\bibfnamefont{A.}~\bibnamefont{Singh}},
		\bibinfo{author}{\bibfnamefont{L.}~\bibnamefont{Harnagea}},
		\bibinfo{author}{\bibfnamefont{A.~K.} \bibnamefont{Sood}}, \bibnamefont{and}
		\bibinfo{author}{\bibfnamefont{A.}~\bibnamefont{Rao}},
		\bibinfo{journal}{Nature communications} \textbf{\bibinfo{volume}{12}},
		\bibinfo{pages}{1699} (\bibinfo{year}{2021}{\natexlab{a}}).
		
		\bibitem[{\citenamefont{Gole{\v{z}} et~al.}(2022)\citenamefont{Gole{\v{z}},
				Dufresne, Kim, Boschini, Chu, Murakami, Levy, Mills, Zhdanovich, Isobe
				et~al.}}]{golevz2022unveiling}
		\bibinfo{author}{\bibfnamefont{D.}~\bibnamefont{Gole{\v{z}}}},
		\bibinfo{author}{\bibfnamefont{S.~K.} \bibnamefont{Dufresne}},
		\bibinfo{author}{\bibfnamefont{M.-J.} \bibnamefont{Kim}},
		\bibinfo{author}{\bibfnamefont{F.}~\bibnamefont{Boschini}},
		\bibinfo{author}{\bibfnamefont{H.}~\bibnamefont{Chu}},
		\bibinfo{author}{\bibfnamefont{Y.}~\bibnamefont{Murakami}},
		\bibinfo{author}{\bibfnamefont{G.}~\bibnamefont{Levy}},
		\bibinfo{author}{\bibfnamefont{A.~K.} \bibnamefont{Mills}},
		\bibinfo{author}{\bibfnamefont{S.}~\bibnamefont{Zhdanovich}},
		\bibinfo{author}{\bibfnamefont{M.}~\bibnamefont{Isobe}},
		\bibnamefont{et~al.}, \bibinfo{journal}{Physical Review B}
		\textbf{\bibinfo{volume}{106}}, \bibinfo{pages}{L121106}
		(\bibinfo{year}{2022}).
		
		\bibitem[{\citenamefont{Baldini et~al.}(2023)\citenamefont{Baldini, Zong, Choi,
				Lee, Michael, Windgaetter, Mazin, Latini, Azoury, Lv
				et~al.}}]{baldini2023spontaneous}
		\bibinfo{author}{\bibfnamefont{E.}~\bibnamefont{Baldini}},
		\bibinfo{author}{\bibfnamefont{A.}~\bibnamefont{Zong}},
		\bibinfo{author}{\bibfnamefont{D.}~\bibnamefont{Choi}},
		\bibinfo{author}{\bibfnamefont{C.}~\bibnamefont{Lee}},
		\bibinfo{author}{\bibfnamefont{M.~H.} \bibnamefont{Michael}},
		\bibinfo{author}{\bibfnamefont{L.}~\bibnamefont{Windgaetter}},
		\bibinfo{author}{\bibfnamefont{I.~I.} \bibnamefont{Mazin}},
		\bibinfo{author}{\bibfnamefont{S.}~\bibnamefont{Latini}},
		\bibinfo{author}{\bibfnamefont{D.}~\bibnamefont{Azoury}},
		\bibinfo{author}{\bibfnamefont{B.}~\bibnamefont{Lv}}, \bibnamefont{et~al.},
		\bibinfo{journal}{Proceedings of the National Academy of Sciences}
		\textbf{\bibinfo{volume}{120}}, \bibinfo{pages}{e2221688120}
		(\bibinfo{year}{2023}).
		
		\bibitem[{\citenamefont{Bretscher
				et~al.}(2021{\natexlab{b}})\citenamefont{Bretscher, Andrich, Murakami,
				Gole{\v{z}}, Remez, Telang, Singh, Harnagea, Cooper, Millis
				et~al.}}]{bretscher2021imaging}
		\bibinfo{author}{\bibfnamefont{H.~M.} \bibnamefont{Bretscher}},
		\bibinfo{author}{\bibfnamefont{P.}~\bibnamefont{Andrich}},
		\bibinfo{author}{\bibfnamefont{Y.}~\bibnamefont{Murakami}},
		\bibinfo{author}{\bibfnamefont{D.}~\bibnamefont{Gole{\v{z}}}},
		\bibinfo{author}{\bibfnamefont{B.}~\bibnamefont{Remez}},
		\bibinfo{author}{\bibfnamefont{P.}~\bibnamefont{Telang}},
		\bibinfo{author}{\bibfnamefont{A.}~\bibnamefont{Singh}},
		\bibinfo{author}{\bibfnamefont{L.}~\bibnamefont{Harnagea}},
		\bibinfo{author}{\bibfnamefont{N.~R.} \bibnamefont{Cooper}},
		\bibinfo{author}{\bibfnamefont{A.~J.} \bibnamefont{Millis}},
		\bibnamefont{et~al.}, \bibinfo{journal}{Science Advances}
		\textbf{\bibinfo{volume}{7}}, \bibinfo{pages}{eabd6147}
		(\bibinfo{year}{2021}{\natexlab{b}}).
		
		\bibitem[{\citenamefont{Chen et~al.}(2022)\citenamefont{Chen, Li, Zhou, Luo,
				Sun, Ye, Sun, Wang, Zheng, Chen et~al.}}]{chen2022optically}
		\bibinfo{author}{\bibfnamefont{F.}~\bibnamefont{Chen}},
		\bibinfo{author}{\bibfnamefont{H.}~\bibnamefont{Li}},
		\bibinfo{author}{\bibfnamefont{H.}~\bibnamefont{Zhou}},
		\bibinfo{author}{\bibfnamefont{S.}~\bibnamefont{Luo}},
		\bibinfo{author}{\bibfnamefont{Z.}~\bibnamefont{Sun}},
		\bibinfo{author}{\bibfnamefont{Z.}~\bibnamefont{Ye}},
		\bibinfo{author}{\bibfnamefont{F.}~\bibnamefont{Sun}},
		\bibinfo{author}{\bibfnamefont{J.}~\bibnamefont{Wang}},
		\bibinfo{author}{\bibfnamefont{Y.}~\bibnamefont{Zheng}},
		\bibinfo{author}{\bibfnamefont{X.}~\bibnamefont{Chen}}, \bibnamefont{et~al.},
		\bibinfo{journal}{Physical Review Letters} \textbf{\bibinfo{volume}{129}},
		\bibinfo{pages}{057402} (\bibinfo{year}{2022}).
		
		\bibitem[{\citenamefont{Eroglu et~al.}(2020)\citenamefont{Eroglu, Comegys,
				Quintanar, Azam, Elafandi, Mahjouri-Samani, and
				Boulesbaa}}]{eroglu2020ultrafast}
		\bibinfo{author}{\bibfnamefont{Z.~E.} \bibnamefont{Eroglu}},
		\bibinfo{author}{\bibfnamefont{O.}~\bibnamefont{Comegys}},
		\bibinfo{author}{\bibfnamefont{L.~S.} \bibnamefont{Quintanar}},
		\bibinfo{author}{\bibfnamefont{N.}~\bibnamefont{Azam}},
		\bibinfo{author}{\bibfnamefont{S.}~\bibnamefont{Elafandi}},
		\bibinfo{author}{\bibfnamefont{M.}~\bibnamefont{Mahjouri-Samani}},
		\bibnamefont{and}
		\bibinfo{author}{\bibfnamefont{A.}~\bibnamefont{Boulesbaa}},
		\bibinfo{journal}{Physical Chemistry Chemical Physics}
		\textbf{\bibinfo{volume}{22}}, \bibinfo{pages}{17385} (\bibinfo{year}{2020}).
		
		\bibitem[{\citenamefont{Mazza et~al.}(2020)\citenamefont{Mazza, R{\"o}sner,
				Windg{\"a}tter, Latini, H{\"u}bener, Millis, Rubio, and
				Georges}}]{mazza2020nature}
		\bibinfo{author}{\bibfnamefont{G.}~\bibnamefont{Mazza}},
		\bibinfo{author}{\bibfnamefont{M.}~\bibnamefont{R{\"o}sner}},
		\bibinfo{author}{\bibfnamefont{L.}~\bibnamefont{Windg{\"a}tter}},
		\bibinfo{author}{\bibfnamefont{S.}~\bibnamefont{Latini}},
		\bibinfo{author}{\bibfnamefont{H.}~\bibnamefont{H{\"u}bener}},
		\bibinfo{author}{\bibfnamefont{A.~J.} \bibnamefont{Millis}},
		\bibinfo{author}{\bibfnamefont{A.}~\bibnamefont{Rubio}}, \bibnamefont{and}
		\bibinfo{author}{\bibfnamefont{A.}~\bibnamefont{Georges}},
		\bibinfo{journal}{Physical Review Letters} \textbf{\bibinfo{volume}{124}},
		\bibinfo{pages}{197601} (\bibinfo{year}{2020}).
		
		\bibitem[{\citenamefont{Zenker et~al.}(2014)\citenamefont{Zenker, Fehske, and
				Beck}}]{zenker2014fate}
		\bibinfo{author}{\bibfnamefont{B.}~\bibnamefont{Zenker}},
		\bibinfo{author}{\bibfnamefont{H.}~\bibnamefont{Fehske}}, \bibnamefont{and}
		\bibinfo{author}{\bibfnamefont{H.}~\bibnamefont{Beck}},
		\bibinfo{journal}{Physical Review B} \textbf{\bibinfo{volume}{90}},
		\bibinfo{pages}{195118} (\bibinfo{year}{2014}).
		
		\bibitem[{\citenamefont{Sun et~al.}(2021)\citenamefont{Sun, Kaneko,
				Gole{\v{z}}, and Millis}}]{sun2021second}
		\bibinfo{author}{\bibfnamefont{Z.}~\bibnamefont{Sun}},
		\bibinfo{author}{\bibfnamefont{T.}~\bibnamefont{Kaneko}},
		\bibinfo{author}{\bibfnamefont{D.}~\bibnamefont{Gole{\v{z}}}},
		\bibnamefont{and} \bibinfo{author}{\bibfnamefont{A.~J.}
			\bibnamefont{Millis}}, \bibinfo{journal}{Physical Review Letters}
		\textbf{\bibinfo{volume}{127}}, \bibinfo{pages}{127702}
		(\bibinfo{year}{2021}).
		
		\bibitem[{\citenamefont{Gole{\v{z}} et~al.}(2020)\citenamefont{Gole{\v{z}},
				Sun, Murakami, Georges, and Millis}}]{golevz2020nonlinear}
		\bibinfo{author}{\bibfnamefont{D.}~\bibnamefont{Gole{\v{z}}}},
		\bibinfo{author}{\bibfnamefont{Z.}~\bibnamefont{Sun}},
		\bibinfo{author}{\bibfnamefont{Y.}~\bibnamefont{Murakami}},
		\bibinfo{author}{\bibfnamefont{A.}~\bibnamefont{Georges}}, \bibnamefont{and}
		\bibinfo{author}{\bibfnamefont{A.~J.} \bibnamefont{Millis}},
		\bibinfo{journal}{Physical Review Letters} \textbf{\bibinfo{volume}{125}},
		\bibinfo{pages}{257601} (\bibinfo{year}{2020}).
		
		\bibitem[{\citenamefont{Khatibi et~al.}(2020)\citenamefont{Khatibi, Ahemeh, and
				Kargarian}}]{khatibi2020excitonic}
		\bibinfo{author}{\bibfnamefont{Z.}~\bibnamefont{Khatibi}},
		\bibinfo{author}{\bibfnamefont{R.}~\bibnamefont{Ahemeh}}, \bibnamefont{and}
		\bibinfo{author}{\bibfnamefont{M.}~\bibnamefont{Kargarian}},
		\bibinfo{journal}{Physical Review B} \textbf{\bibinfo{volume}{102}},
		\bibinfo{pages}{245121} (\bibinfo{year}{2020}).
		
		\bibitem[{\citenamefont{Murakami et~al.}(2020)\citenamefont{Murakami,
				Gole{\v{z}}, Kaneko, Koga, Millis, and Werner}}]{murakami2020collective}
		\bibinfo{author}{\bibfnamefont{Y.}~\bibnamefont{Murakami}},
		\bibinfo{author}{\bibfnamefont{D.}~\bibnamefont{Gole{\v{z}}}},
		\bibinfo{author}{\bibfnamefont{T.}~\bibnamefont{Kaneko}},
		\bibinfo{author}{\bibfnamefont{A.}~\bibnamefont{Koga}},
		\bibinfo{author}{\bibfnamefont{A.~J.} \bibnamefont{Millis}},
		\bibnamefont{and} \bibinfo{author}{\bibfnamefont{P.}~\bibnamefont{Werner}},
		\bibinfo{journal}{Physical Review B} \textbf{\bibinfo{volume}{101}},
		\bibinfo{pages}{195118} (\bibinfo{year}{2020}).
		
		\bibitem[{\citenamefont{Moon}(2021)}]{moon2021metal}
		\bibinfo{author}{\bibfnamefont{B.~H.} \bibnamefont{Moon}},
		\bibinfo{journal}{Emergent Materials} \textbf{\bibinfo{volume}{4}},
		\bibinfo{pages}{989} (\bibinfo{year}{2021}).
		
		\bibitem[{\citenamefont{Thilagam}(2014)}]{thilagam2014exciton}
		\bibinfo{author}{\bibfnamefont{A.}~\bibnamefont{Thilagam}},
		\bibinfo{journal}{Journal of Applied Physics} \textbf{\bibinfo{volume}{116}}
		(\bibinfo{year}{2014}).
		
		\bibitem[{\citenamefont{Nag et~al.}(2019)\citenamefont{Nag, Slager, Higuchi,
				and Oka}}]{nag2019dynamical}
		\bibinfo{author}{\bibfnamefont{T.}~\bibnamefont{Nag}},
		\bibinfo{author}{\bibfnamefont{R.-J.} \bibnamefont{Slager}},
		\bibinfo{author}{\bibfnamefont{T.}~\bibnamefont{Higuchi}}, \bibnamefont{and}
		\bibinfo{author}{\bibfnamefont{T.}~\bibnamefont{Oka}},
		\bibinfo{journal}{Physical Review B} \textbf{\bibinfo{volume}{100}},
		\bibinfo{pages}{134301} (\bibinfo{year}{2019}).
		
		\bibitem[{\citenamefont{Rodrigues et~al.}(2016)\citenamefont{Rodrigues, Peron,
				Ji, and Kurths}}]{2016kuramoto}
		\bibinfo{author}{\bibfnamefont{F.~A.} \bibnamefont{Rodrigues}},
		\bibinfo{author}{\bibfnamefont{T.~K.~D.} \bibnamefont{Peron}},
		\bibinfo{author}{\bibfnamefont{P.}~\bibnamefont{Ji}}, \bibnamefont{and}
		\bibinfo{author}{\bibfnamefont{J.}~\bibnamefont{Kurths}},
		\bibinfo{journal}{Physics Reports} \textbf{\bibinfo{volume}{610}},
		\bibinfo{pages}{1} (\bibinfo{year}{2016}).
		
		\bibitem[{\citenamefont{Kuramoto}(1975)}]{kuramoto1975international}
		\bibinfo{author}{\bibfnamefont{Y.}~\bibnamefont{Kuramoto}},
		\bibinfo{journal}{Lecture notes in Physics} \textbf{\bibinfo{volume}{30}},
		\bibinfo{pages}{420} (\bibinfo{year}{1975}).
		
		\bibitem[{\citenamefont{Acebr{\'o}n et~al.}(2005)\citenamefont{Acebr{\'o}n,
				Bonilla, Vicente, Ritort, and Spigler}}]{acebron2005kuramoto}
		\bibinfo{author}{\bibfnamefont{J.~A.} \bibnamefont{Acebr{\'o}n}},
		\bibinfo{author}{\bibfnamefont{L.~L.} \bibnamefont{Bonilla}},
		\bibinfo{author}{\bibfnamefont{C.~J.~P.} \bibnamefont{Vicente}},
		\bibinfo{author}{\bibfnamefont{F.}~\bibnamefont{Ritort}}, \bibnamefont{and}
		\bibinfo{author}{\bibfnamefont{R.}~\bibnamefont{Spigler}},
		\bibinfo{journal}{Reviews of modern physics} \textbf{\bibinfo{volume}{77}},
		\bibinfo{pages}{137} (\bibinfo{year}{2005}).
		
		\bibitem[{\citenamefont{Lotfi et~al.}(2018)\citenamefont{Lotfi, Rodrigues, and
				Darooneh}}]{lotfi2018role}
		\bibinfo{author}{\bibfnamefont{N.}~\bibnamefont{Lotfi}},
		\bibinfo{author}{\bibfnamefont{F.~A.} \bibnamefont{Rodrigues}},
		\bibnamefont{and} \bibinfo{author}{\bibfnamefont{A.~H.}
			\bibnamefont{Darooneh}}, \bibinfo{journal}{Chaos: An Interdisciplinary
			Journal of Nonlinear Science} \textbf{\bibinfo{volume}{28}},
		\bibinfo{pages}{033102} (\bibinfo{year}{2018}).
		
		\bibitem[{\citenamefont{Strogatz et~al.}(1988)\citenamefont{Strogatz, Marcus,
				Westervelt, and Mirollo}}]{1988simple}
		\bibinfo{author}{\bibfnamefont{S.}~\bibnamefont{Strogatz}},
		\bibinfo{author}{\bibfnamefont{C.}~\bibnamefont{Marcus}},
		\bibinfo{author}{\bibfnamefont{R.}~\bibnamefont{Westervelt}},
		\bibnamefont{and} \bibinfo{author}{\bibfnamefont{R.}~\bibnamefont{Mirollo}},
		\bibinfo{journal}{Physical review letters} \textbf{\bibinfo{volume}{61}},
		\bibinfo{pages}{2380} (\bibinfo{year}{1988}).
		
		\bibitem[{\citenamefont{Moreira and de~Aguiar}(2019)}]{moreira2019global}
		\bibinfo{author}{\bibfnamefont{C.~A.} \bibnamefont{Moreira}} \bibnamefont{and}
		\bibinfo{author}{\bibfnamefont{M.~A.} \bibnamefont{de~Aguiar}},
		\bibinfo{journal}{Physica A: Statistical Mechanics and its Applications}
		\textbf{\bibinfo{volume}{514}}, \bibinfo{pages}{487} (\bibinfo{year}{2019}).
		
		\bibitem[{\citenamefont{Anderson}(1958)}]{Anderson}
		\bibinfo{author}{\bibfnamefont{P.~W.} \bibnamefont{Anderson}},
		\bibinfo{journal}{Physical Review} \textbf{\bibinfo{volume}{112}},
		\bibinfo{pages}{1900} (\bibinfo{year}{1958}).
		
		\bibitem[{\citenamefont{Murakami et~al.}(2017)\citenamefont{Murakami,
				Gole{\v{z}}, Eckstein, and Werner}}]{PWerner}
		\bibinfo{author}{\bibfnamefont{Y.}~\bibnamefont{Murakami}},
		\bibinfo{author}{\bibfnamefont{D.}~\bibnamefont{Gole{\v{z}}}},
		\bibinfo{author}{\bibfnamefont{M.}~\bibnamefont{Eckstein}}, \bibnamefont{and}
		\bibinfo{author}{\bibfnamefont{P.}~\bibnamefont{Werner}},
		\bibinfo{journal}{Physical Review Letters} \textbf{\bibinfo{volume}{119}},
		\bibinfo{pages}{247601} (\bibinfo{year}{2017}).
		
		\bibitem[{\citenamefont{Goldman et~al.}(2014)\citenamefont{Goldman,
				Juzeli{\=u}nas, {\"O}hberg, and Spielman}}]{FloquetGoldman}
		\bibinfo{author}{\bibfnamefont{N.}~\bibnamefont{Goldman}},
		\bibinfo{author}{\bibfnamefont{G.}~\bibnamefont{Juzeli{\=u}nas}},
		\bibinfo{author}{\bibfnamefont{P.}~\bibnamefont{{\"O}hberg}},
		\bibnamefont{and} \bibinfo{author}{\bibfnamefont{I.~B.}
			\bibnamefont{Spielman}}, \bibinfo{journal}{Reports on Progress in Physics}
		\textbf{\bibinfo{volume}{77}}, \bibinfo{pages}{126401}
		(\bibinfo{year}{2014}).
		
		\bibitem[{\citenamefont{Meinert et~al.}(2016)\citenamefont{Meinert, Mark,
				Lauber, Daley, and N{\"a}gerl}}]{meinert2016floquet}
		\bibinfo{author}{\bibfnamefont{F.}~\bibnamefont{Meinert}},
		\bibinfo{author}{\bibfnamefont{M.~J.} \bibnamefont{Mark}},
		\bibinfo{author}{\bibfnamefont{K.}~\bibnamefont{Lauber}},
		\bibinfo{author}{\bibfnamefont{A.~J.} \bibnamefont{Daley}}, \bibnamefont{and}
		\bibinfo{author}{\bibfnamefont{H.-C.} \bibnamefont{N{\"a}gerl}},
		\bibinfo{journal}{Physical review letters} \textbf{\bibinfo{volume}{116}},
		\bibinfo{pages}{205301} (\bibinfo{year}{2016}).
		
		\bibitem[{\citenamefont{Schweizer et~al.}(2019)\citenamefont{Schweizer, Grusdt,
				Berngruber, Barbiero, Demler, Goldman, Bloch, and
				Aidelsburger}}]{schweizer2019floquet}
		\bibinfo{author}{\bibfnamefont{C.}~\bibnamefont{Schweizer}},
		\bibinfo{author}{\bibfnamefont{F.}~\bibnamefont{Grusdt}},
		\bibinfo{author}{\bibfnamefont{M.}~\bibnamefont{Berngruber}},
		\bibinfo{author}{\bibfnamefont{L.}~\bibnamefont{Barbiero}},
		\bibinfo{author}{\bibfnamefont{E.}~\bibnamefont{Demler}},
		\bibinfo{author}{\bibfnamefont{N.}~\bibnamefont{Goldman}},
		\bibinfo{author}{\bibfnamefont{I.}~\bibnamefont{Bloch}}, \bibnamefont{and}
		\bibinfo{author}{\bibfnamefont{M.}~\bibnamefont{Aidelsburger}},
		\bibinfo{journal}{Nature Physics} \textbf{\bibinfo{volume}{15}},
		\bibinfo{pages}{1168} (\bibinfo{year}{2019}).
		
		\bibitem[{\citenamefont{Eckardt}(2017)}]{Floquetreview}
		\bibinfo{author}{\bibfnamefont{A.}~\bibnamefont{Eckardt}},
		\bibinfo{journal}{Reviews of Modern Physics} \textbf{\bibinfo{volume}{89}},
		\bibinfo{pages}{011004} (\bibinfo{year}{2017}).
		
		\bibitem[{\citenamefont{Oka and Kitamura}(2019)}]{FloquetOka}
		\bibinfo{author}{\bibfnamefont{T.}~\bibnamefont{Oka}} \bibnamefont{and}
		\bibinfo{author}{\bibfnamefont{S.}~\bibnamefont{Kitamura}},
		\bibinfo{journal}{Annual Review of Condensed Matter Physics}
		\textbf{\bibinfo{volume}{10}}, \bibinfo{pages}{387} (\bibinfo{year}{2019}).
		
		\bibitem[{\citenamefont{Schuster et~al.}(2021)\citenamefont{Schuster, Flicker,
				Li, Kotochigova, Moore, Ye, and Yao}}]{schuster2021floquet}
		\bibinfo{author}{\bibfnamefont{T.}~\bibnamefont{Schuster}},
		\bibinfo{author}{\bibfnamefont{F.}~\bibnamefont{Flicker}},
		\bibinfo{author}{\bibfnamefont{M.}~\bibnamefont{Li}},
		\bibinfo{author}{\bibfnamefont{S.}~\bibnamefont{Kotochigova}},
		\bibinfo{author}{\bibfnamefont{J.~E.} \bibnamefont{Moore}},
		\bibinfo{author}{\bibfnamefont{J.}~\bibnamefont{Ye}}, \bibnamefont{and}
		\bibinfo{author}{\bibfnamefont{N.~Y.} \bibnamefont{Yao}},
		\bibinfo{journal}{Physical Review A} \textbf{\bibinfo{volume}{103}},
		\bibinfo{pages}{063322} (\bibinfo{year}{2021}).
		
		\bibitem[{\citenamefont{Mikami et~al.}(2016)\citenamefont{Mikami, Kitamura,
				Yasuda, Tsuji, Oka, and Aoki}}]{mikami2016brillouin}
		\bibinfo{author}{\bibfnamefont{T.}~\bibnamefont{Mikami}},
		\bibinfo{author}{\bibfnamefont{S.}~\bibnamefont{Kitamura}},
		\bibinfo{author}{\bibfnamefont{K.}~\bibnamefont{Yasuda}},
		\bibinfo{author}{\bibfnamefont{N.}~\bibnamefont{Tsuji}},
		\bibinfo{author}{\bibfnamefont{T.}~\bibnamefont{Oka}}, \bibnamefont{and}
		\bibinfo{author}{\bibfnamefont{H.}~\bibnamefont{Aoki}},
		\bibinfo{journal}{Physical Review B} \textbf{\bibinfo{volume}{93}},
		\bibinfo{pages}{144307} (\bibinfo{year}{2016}).
		
		\bibitem[{\citenamefont{Trevisan et~al.}(2022)\citenamefont{Trevisan,
				Villar~Arribi, Heinonen, Slager, and Orth}}]{trevisan2022bicircular}
		\bibinfo{author}{\bibfnamefont{T.~V.} \bibnamefont{Trevisan}},
		\bibinfo{author}{\bibfnamefont{P.}~\bibnamefont{Villar~Arribi}},
		\bibinfo{author}{\bibfnamefont{O.}~\bibnamefont{Heinonen}},
		\bibinfo{author}{\bibfnamefont{R.-J.} \bibnamefont{Slager}},
		\bibnamefont{and} \bibinfo{author}{\bibfnamefont{P.}~\bibnamefont{Orth}}
		(\bibinfo{year}{2022}).
		
		\bibitem[{\citenamefont{Kuramoto and Kuramoto}(1984)}]{kuramoto1984chemical}
		\bibinfo{author}{\bibfnamefont{Y.}~\bibnamefont{Kuramoto}} \bibnamefont{and}
		\bibinfo{author}{\bibfnamefont{Y.}~\bibnamefont{Kuramoto}},
		\emph{\bibinfo{title}{Chemical turbulence}} (\bibinfo{publisher}{Springer},
		\bibinfo{year}{1984}).
		
		\bibitem[{\citenamefont{Kaneko et~al.}(2013)\citenamefont{Kaneko, Toriyama,
				Konishi, and Ohta}}]{kaneko2013orthorhombic}
		\bibinfo{author}{\bibfnamefont{T.}~\bibnamefont{Kaneko}},
		\bibinfo{author}{\bibfnamefont{T.}~\bibnamefont{Toriyama}},
		\bibinfo{author}{\bibfnamefont{T.}~\bibnamefont{Konishi}}, \bibnamefont{and}
		\bibinfo{author}{\bibfnamefont{Y.}~\bibnamefont{Ohta}},
		\bibinfo{journal}{Physical Review B} \textbf{\bibinfo{volume}{87}},
		\bibinfo{pages}{035121} (\bibinfo{year}{2013}).
		
		\bibitem[{\citenamefont{Yusupov et~al.}(2010)\citenamefont{Yusupov, Mertelj,
				Kabanov, Brazovskii, Kusar, Chu, Fisher, and
				Mihailovic}}]{yusupov2010coherent}
		\bibinfo{author}{\bibfnamefont{R.}~\bibnamefont{Yusupov}},
		\bibinfo{author}{\bibfnamefont{T.}~\bibnamefont{Mertelj}},
		\bibinfo{author}{\bibfnamefont{V.~V.} \bibnamefont{Kabanov}},
		\bibinfo{author}{\bibfnamefont{S.}~\bibnamefont{Brazovskii}},
		\bibinfo{author}{\bibfnamefont{P.}~\bibnamefont{Kusar}},
		\bibinfo{author}{\bibfnamefont{J.-H.} \bibnamefont{Chu}},
		\bibinfo{author}{\bibfnamefont{I.~R.} \bibnamefont{Fisher}},
		\bibnamefont{and}
		\bibinfo{author}{\bibfnamefont{D.}~\bibnamefont{Mihailovic}},
		\bibinfo{journal}{Nature Physics} \textbf{\bibinfo{volume}{6}},
		\bibinfo{pages}{681} (\bibinfo{year}{2010}).
		
		\bibitem[{\citenamefont{Matsunaga et~al.}(2017)\citenamefont{Matsunaga, Tsuji,
				Makise, Terai, Aoki, and Shimano}}]{matsunaga2017polarization}
		\bibinfo{author}{\bibfnamefont{R.}~\bibnamefont{Matsunaga}},
		\bibinfo{author}{\bibfnamefont{N.}~\bibnamefont{Tsuji}},
		\bibinfo{author}{\bibfnamefont{K.}~\bibnamefont{Makise}},
		\bibinfo{author}{\bibfnamefont{H.}~\bibnamefont{Terai}},
		\bibinfo{author}{\bibfnamefont{H.}~\bibnamefont{Aoki}}, \bibnamefont{and}
		\bibinfo{author}{\bibfnamefont{R.}~\bibnamefont{Shimano}},
		\bibinfo{journal}{arXiv preprint arXiv:1703.02815}  (\bibinfo{year}{2017}).
		
		\bibitem[{\citenamefont{Sun and Millis}(2020)}]{sun2020bardasis}
		\bibinfo{author}{\bibfnamefont{Z.}~\bibnamefont{Sun}} \bibnamefont{and}
		\bibinfo{author}{\bibfnamefont{A.~J.} \bibnamefont{Millis}},
		\bibinfo{journal}{Physical Review B} \textbf{\bibinfo{volume}{102}},
		\bibinfo{pages}{041110} (\bibinfo{year}{2020}).
		
		\bibitem[{\citenamefont{Shirley}(1965)}]{FloquetSolution}
		\bibinfo{author}{\bibfnamefont{J.~H.} \bibnamefont{Shirley}},
		\bibinfo{journal}{Physical Review} \textbf{\bibinfo{volume}{138}},
		\bibinfo{pages}{B979} (\bibinfo{year}{1965}).
		
		\bibitem[{\citenamefont{Autler and Townes}(1955)}]{autler1955stark}
		\bibinfo{author}{\bibfnamefont{S.~H.} \bibnamefont{Autler}} \bibnamefont{and}
		\bibinfo{author}{\bibfnamefont{C.~H.} \bibnamefont{Townes}},
		\bibinfo{journal}{Physical Review} \textbf{\bibinfo{volume}{100}},
		\bibinfo{pages}{703} (\bibinfo{year}{1955}).
		
		\bibitem[{\citenamefont{Zel'Dovich}(1967)}]{zel1967quasienergy}
		\bibinfo{author}{\bibfnamefont{Y.~B.} \bibnamefont{Zel'Dovich}},
		\bibinfo{journal}{Sov. Phys. JETP} \textbf{\bibinfo{volume}{24}}
		(\bibinfo{year}{1967}).
		
		\bibitem[{\citenamefont{Sambe}(1973)}]{sambe1973steady}
		\bibinfo{author}{\bibfnamefont{H.}~\bibnamefont{Sambe}},
		\bibinfo{journal}{Physical Review A} \textbf{\bibinfo{volume}{7}},
		\bibinfo{pages}{2203} (\bibinfo{year}{1973}).
		
		\bibitem[{\citenamefont{Rahav et~al.}(2003)\citenamefont{Rahav, Gilary, and
				Fishman}}]{rahav2003effective}
		\bibinfo{author}{\bibfnamefont{S.}~\bibnamefont{Rahav}},
		\bibinfo{author}{\bibfnamefont{I.}~\bibnamefont{Gilary}}, \bibnamefont{and}
		\bibinfo{author}{\bibfnamefont{S.}~\bibnamefont{Fishman}},
		\bibinfo{journal}{Physical Review A} \textbf{\bibinfo{volume}{68}},
		\bibinfo{pages}{013820} (\bibinfo{year}{2003}).
		
		\bibitem[{\citenamefont{Casas et~al.}(2001)\citenamefont{Casas, Oteo, and
				Ros}}]{casas2001floquet}
		\bibinfo{author}{\bibfnamefont{F.}~\bibnamefont{Casas}},
		\bibinfo{author}{\bibfnamefont{J.}~\bibnamefont{Oteo}}, \bibnamefont{and}
		\bibinfo{author}{\bibfnamefont{J.}~\bibnamefont{Ros}},
		\bibinfo{journal}{Journal of Physics A: Mathematical and General}
		\textbf{\bibinfo{volume}{34}}, \bibinfo{pages}{3379} (\bibinfo{year}{2001}).
		
		\bibitem[{\citenamefont{Blanes et~al.}(2009)\citenamefont{Blanes, Casas, Oteo,
				and Ros}}]{blanes2009magnus}
		\bibinfo{author}{\bibfnamefont{S.}~\bibnamefont{Blanes}},
		\bibinfo{author}{\bibfnamefont{F.}~\bibnamefont{Casas}},
		\bibinfo{author}{\bibfnamefont{J.-A.} \bibnamefont{Oteo}}, \bibnamefont{and}
		\bibinfo{author}{\bibfnamefont{J.}~\bibnamefont{Ros}},
		\bibinfo{journal}{Physics reports} \textbf{\bibinfo{volume}{470}},
		\bibinfo{pages}{151} (\bibinfo{year}{2009}).
		
		\bibitem[{\citenamefont{Mananga and
				Charpentier}(2011)}]{mananga2011introduction}
		\bibinfo{author}{\bibfnamefont{E.~S.} \bibnamefont{Mananga}} \bibnamefont{and}
		\bibinfo{author}{\bibfnamefont{T.}~\bibnamefont{Charpentier}},
		\bibinfo{journal}{The Journal of chemical physics}
		\textbf{\bibinfo{volume}{135}}, \bibinfo{pages}{044109}
		(\bibinfo{year}{2011}).
		
		\bibitem[{\citenamefont{Press and Teukolsky}(1992)}]{press1992adaptive}
		\bibinfo{author}{\bibfnamefont{W.~H.} \bibnamefont{Press}} \bibnamefont{and}
		\bibinfo{author}{\bibfnamefont{S.~A.} \bibnamefont{Teukolsky}},
		\bibinfo{journal}{Computers in Physics} \textbf{\bibinfo{volume}{6}},
		\bibinfo{pages}{188} (\bibinfo{year}{1992}).
		
	\end{thebibliography}

\end{document}